\documentclass{IEEEtran}

\newif\ifreview

\reviewfalse

\usepackage[]{graphicx}

\usepackage{cite}
\usepackage{amsmath,amssymb,amsfonts}
\usepackage{textcomp}
\usepackage[normalem]{ulem}
\usepackage[hidelinks]{hyperref}
\usepackage[caption=false]{subfig} 

\newcommand{\curl}{\nabla \times}

\def\BibTeX{{\rm B\kern-.05em{\sc i\kern-.025em b}\kern-.08em
    T\kern-.1667em\lower.7ex\hbox{E}\kern-.125emX}}

\begin{document}

\title{\vspace{-0.3in}A High-Order-Accurate 3D Surface Integral Equation Solver for Uniaxial Anisotropic Media}

\author{Jin Hu, \IEEEmembership{Student Member, IEEE} and Constantine Sideris, \IEEEmembership{Member, IEEE}

\thanks{The authors gratefully acknowledge support by the Air Force Office of Scientific Research (FA9550-20-1-0087) and the National Science Foundation (CCF-1849965, CCF-2047433).}

\thanks{J. Hu and C. Sideris are with the Department of Electrical and Computer Engineering, University of Southern California, Los Angeles, CA 90089, USA (e-mails: jinhu@usc.edu, csideris@usc.edu).}}



\maketitle

\begin{abstract}
This paper introduces a high-order accurate surface integral equation method for solving 3D electromagnetic scattering for dielectric objects with uniaxially anisotropic permittivity tensors. The N-M\"uller formulation is leveraged resulting in a second-kind integral formulation, and a finite-difference-based approach is used to deal with the strongly singular terms resulting from the dyadic Green's functions for uniaxially anisotropic media while maintaining the high-order accuracy of the discretization strategy. The integral operators are discretized via a Nystr\"om-collocation approach, which represents the unknown surface densities in terms of Chebyshev polynomials on curvilinear quadrilateral surface patches. The convergence is investigated for various geometries, including a sphere, cube, a complicated NURBS geometry imported from a 3D CAD modeler software, and a nanophotonic silicon waveguide, and results are compared against a commercial finite element solver. To the best of our knowledge, this is the first demonstration of high-order accuracy for objects with uniaxially anisotropic materials using surface integral equations.
\end{abstract}


\section{Introduction\label{sec:introduction}}
Boundary Integral Equations (BIE) are a powerful approach for numerically solving Maxwell's equations and have been applied to solve a plethora of scattering problems, including antennas~\cite{makarov2001mom}, radar scattering~\cite{yla2005application}, and most recently nanophotonics~\cite{sideris2019ultrafast,garza2021high,garza2021boundary}. Traditionally known as open boundary problems due to satisfying the Sommerfeld radiation condition by design, BIE's have also recently been successfully applied for solving dielectric waveguiding problems, which require simulating waveguides extending to and from infinity, in both two~\cite{sideris2019ultrafast,bruno2017windowed} and three~\cite{garza2021boundary} dimensions. Unlike other volumetric computational approaches, such as Finite Difference (FD) and Finite Element (FE) methods, which require generating complicated volume meshes, BIE methods only mesh the surfaces between material regions. Since BIE methods solve for unknowns over surface rather than volume meshes, they may also result in significantly smaller problems compared to using volumetric approaches in scenarios with high volume to surface area ratios. BIE's have predominantly been used for solving problems with homogeneous, isotropic dielectrics due to the availability of closed-form Dyadic Green's functions, which can be readily discretized using suitable numerical quadrature and singularity treatment approaches. For example, our recent work in~\cite{hu2021chebyshev} demonstrates high-order convergence discretizing the Magnetic Field Integral Equation (MFIE) and the N-M\"uller formulation for modeling metals and dielectrics respectively using a Chebyshev-based Nystr\"om method. On the other hand, many anisotropic materials are commonly used in engineering applications, such as anisotropic dielectric substrates for antennas~\cite{pozar1987radiation, mumcu2005rf} and liquid crystal claddings for designing reconfigurable nanophotonic devices~\cite{pfeifle2012silicon}. However, despite the fact that closed-form Green's functions have been derived for uniaxially anisotropic media, there is a dearth of work available using BIE methods to solve problems with these materials. In fact, the only discretization approach in the literature is~\cite{mumcu2008surface}, which presents compelling results comparing against volumetric methods but does not report on error or convergence properties.

Indeed, although closed form expressions for the dyadic Green's functions for materials with uniaxially anisotropic permittivity and permeability do exist~\cite{weiglhofer1990dyadic}, they are significantly more complex and challenging to discretize than the corresponding expressions for the isotropic material case (e.g., see eq.~\ref{eq:Ddyadic}). The PMCHWT~\cite{poggio1970integral} formulation is used in~\cite{mumcu2008surface} and discretized using the Method-of-Moments (MoM) and RWG basis functions~\cite{rao1982electromagnetic,yla2003calculation}. The strongly singular part of the $\overline{\mathbf{G}}_{ee}$ operator (known as the $T$ operator in the literature for the isotropic case) is dealt with in the usual manner by using integration by parts to decrease the kernel singularity by moving a derivative to the testing function. However, the $\overline{\mathbf{G}}_{em} \propto \left(\curl \overline{\mathbf{G}}_{ee}\right)$ operator (known as the $K$ operator in the literature for the isotropic case) also contains a strong singularity which cannot be easily reduced. \cite{mumcu2008surface} approximates integrals with $\overline{\mathbf{G}}_{em}$ by shifting the target point $\mathbf{r}$ slightly off the surface. Unfortunately, this approach is expected to result in poor accuracy since the operator is evaluated on a different target point than the original intended one on the surface, and furthermore because the kernel remains nearly singular and is therefore very challenging to numerically integrate even with the target point being shifted off the surface.

In this work, we present an new discretization strategy, which when combined with the singular integration approach using Chebhyshev polynomials to represent the unknown densities introduced in~\cite{hu2021chebyshev} achieves high-order accuracy for scattering from objects composed of uniaxially anisotropic materials. To the best of our knowledge, this is the very first demonstration of a boundary integral solver for anisotropic media which achieves high-order accuracy. Note that in all of our examples we assume that only the permittivity tensor is anisotropic and that $\mu_r=1$; however, the approach presented can readily be extended to support materials with both permittivity and permeability tensors having anisotropy. The paper is organized as follows. Section II briefly introduces the surface integral formulation under consideration for dielectric scatterers. Section III reviews the Dyadic Green's functions for uniaxial anisotropic media and sets up a system of integral equations for a scenario with an anisotropic scatterer inside an isotropic exterior medium based on the N-M\"uller formulation. Section IV analyzes the singular behavior of each anisotropic kernel operator. Section V presents our Chebyshev-based discretization and singular integration approach for accurate evaluation of the integral operators. Finally, Section VI demonstrates error convergence and both near and far-field numerical results for four different example cases.

\section{Surface Integral Equation Formulation\label{sec:formulation}}
\begin{figure}[t]
    \centering
\includegraphics[width=\columnwidth]{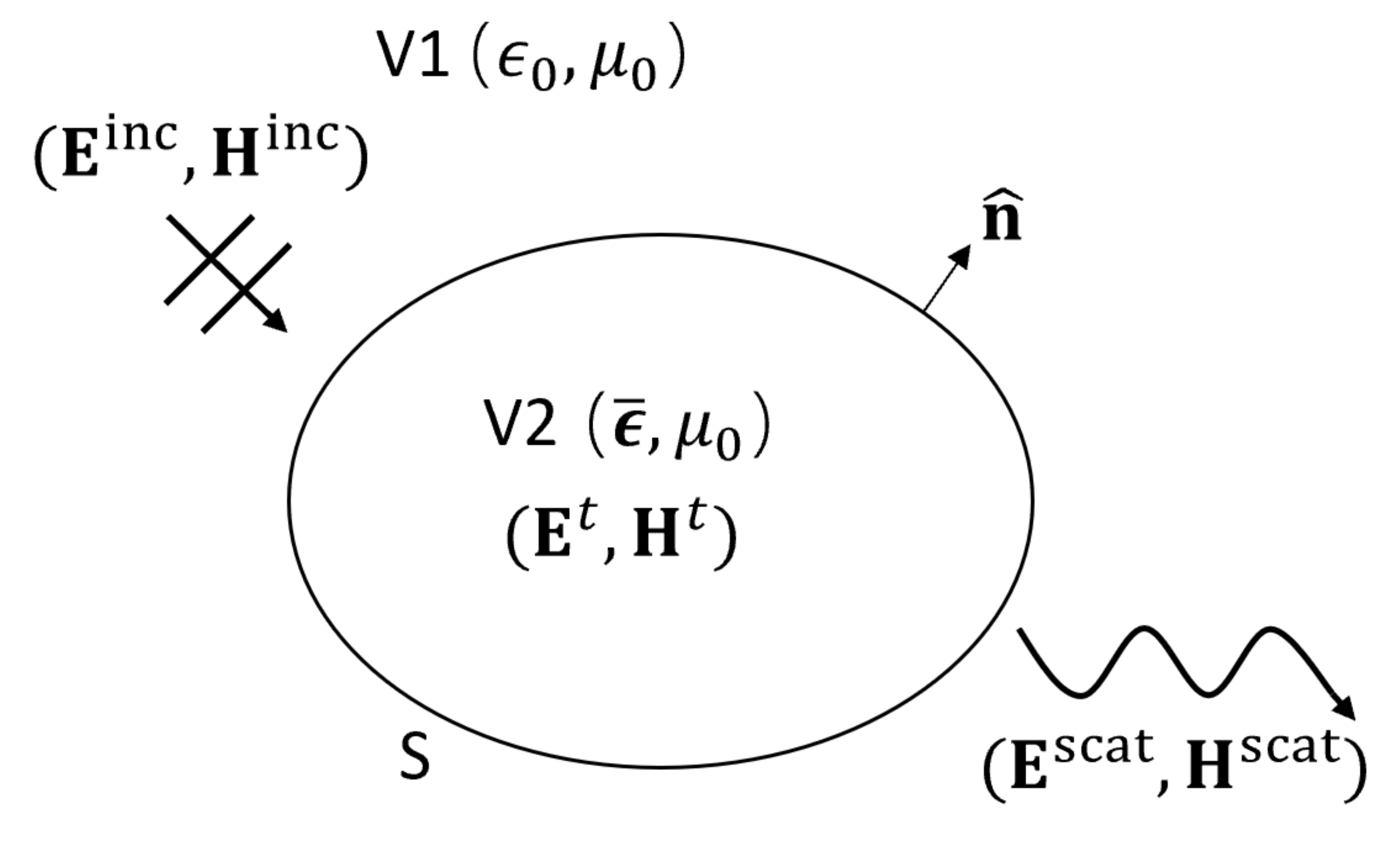}
\caption{Electromagnetic scattering from a uniaxial anisotropic object illuminated by an incident wave excitation}
\label{fig:formulation}
\end{figure}

We consider the problem of evaluating the scattered field from a non-magnetic uniaxial anisotropic object ($V_{2}$) embedded in a free space region ($V_{1}$) as shown in Fig.~\ref{fig:formulation}. Note that for the subsequent derivations we assume $V_1$ is free-space without loss of generality; however, it can also be any arbitrary isotropic homogeneous background medium. The object is illuminated by an incident field excitation $\left(\mathbf{E}^\text{inc},\mathbf{H}^\text{inc}\right)$ that will lead to both scattered fields $\left(\mathbf{E}^\text{scat},\mathbf{H}^\text{scat}\right)$ outside the object and transmitted fields $\left(\mathbf{E}^\text{t},\mathbf{H}^\text{t}\right)$ inside the object. 

To obtain an equivalent problem for the exterior region based on surface equivalence principle, the interior fields can be nulled, and the total fields in the exterior region $\left(\mathbf{E}_{1},\mathbf{H}_{1}\right)$ are a superposition of incident and scattered fields, which can be represented as:
\begin{equation}
    \mathbf{E}_{1} = \mathbf{E}^\text{inc} + \int_{S} \overline{\mathbf{G}}_{em}^{1}\cdot\mathbf{M}_{1}d\sigma(\mathbf{r'}) + \int_{S} \overline{\mathbf{G}}_{ee}^{1}\cdot\mathbf{J}_{1}d\sigma(\mathbf{r'})  
    \label{eq:representationE1}
\end{equation}
\begin{equation}
      \mathbf{H}_{1} = \mathbf{H}^\text{inc} + \int_{S} \overline{\mathbf{G}}_{mm}^{1}\cdot\mathbf{M}_{1}d\sigma(\mathbf{r'}) + \int_{S} \overline{\mathbf{G}}_{me}^{1}\cdot\mathbf{J}_{1}d\sigma(\mathbf{r'})
      \label{eq:representationH1}
\end{equation}
$\mathbf{J}_{1} = \mathbf{\hat{n}} \times \mathbf{H}_{1}$ and $\mathbf{M}_{1} = \mathbf{E}_{1} \times \mathbf{\hat{n}} $ are the equivalent surface electric and magnetic current densities for the exterior region. $\overline{\mathbf{G}}_{ee}^{1}$ and $\overline{\mathbf{G}}_{em}^{1}$ (resp. $\overline{\mathbf{G}}_{me}^{1}$ and $\overline{\mathbf{G}}_{mm}^{1}$) are the dyadic Green's functions of the exterior region, corresponding to the electric fields (resp. magnetic fields) produced by delta electric and magnetic current sources respectively in $V_{1}$. By letting the target point $\mathbf{r}$ approach the surface S from the exterior $V_{1}$ and taking the cross products of eqs. \eqref{eq:representationE1} and \eqref{eq:representationH1} with the unit normal vector to the surface $\mathbf{\hat{n}}$, the first set of equations is obtained as
\begin{equation}
\frac{1}{2}\mathbf{M}_{1} + \mathcal{K}^{1}_{em}\mathbf{M}_{1} +
\mathcal{K}^{1}_{ee}\mathbf{J}_{1} = -\mathbf{\hat{n}} \times \mathbf{E}^\text{inc} 
\label{eq:EFIE1}
\end{equation}
\begin{equation}
\frac{1}{2}\mathbf{J}_{1} - \mathcal{K}^{1}_{mm}\mathbf{M}_{1} -
\mathcal{K}^{1}_{me}\mathbf{J}_{1} = \mathbf{\hat{n}} \times \mathbf{H}^\text{inc}
\label{eq:MFIE1}
\end{equation}
with 
\begin{equation}
    \mathcal{K}^{1}_{\alpha\beta}[\mathbf{a}](\mathbf{r}) = 
    \mathbf{\hat{n}}(\mathbf{r}) \times \int_{S} \overline{\mathbf{G}}_{\alpha\beta}^{1}(\mathbf{r},\mathbf{r}')\cdot\mathbf{a}(\mathbf{r}')d\sigma(\mathbf{r'}) \quad \mathbf{r} \in S
     \label{eq:operatorK1} 
\end{equation}
where the subscripts $\alpha$ and $\beta$ can be either $e$ or $m$. 
Similarly, the equivalent problem for the interior region can be formulated by setting the exterior field to zero, allowing the total fields inside the anisotropic uniaxial region $\left(\mathbf{E}_{2},\mathbf{H}_{2}\right)$, which are the same as the transmitted fields, to be represented as: 
\begin{equation}
    \mathbf{E}_{2} = \int_{S} \overline{\mathbf{G}}_{em}^{2}\cdot\mathbf{M}_{2}d\sigma(\mathbf{r'}) + \int_{S} \overline{\mathbf{G}}_{ee}^{2}\cdot\mathbf{J}_{2}d\sigma(\mathbf{r'}) 
    \label{eq:representationE2}
\end{equation}
\begin{equation}
     \mathbf{H}_{2} = \int_{S} \overline{\mathbf{G}}_{mm}^{2}\cdot\mathbf{M}_{2}d\sigma(\mathbf{r'}) + \int_{S} \overline{\mathbf{G}}_{me}^{2}\cdot\mathbf{J}_{2}d\sigma(\mathbf{r'})
     \label{eq:representationH2}
\end{equation}
Analogous to the exterior problem, $\mathbf{J}_{2} = (-\mathbf{\hat{n}}) \times \mathbf{H}_{2}$ and $\mathbf{M}_{2} = \mathbf{E}_{2} \times (-\mathbf{\hat{n}}) $ are the equivalent surface electric and magnetic current densities for the interior problem, $\overline{\mathbf{G}}_{ee}^{2}$, $\overline{\mathbf{G}}_{em}^{2}$,
$\overline{\mathbf{G}}_{me}^{2}$ and 
$\overline{\mathbf{G}}_{mm}^{2}$ are the four dyadic Green's functions for the internal anisotropic uniaxial region $V_{2}$ for which the subscripts can be interpreted in the same manner as the exterior functions described above. By letting the target point $\mathbf{r}$ approach the surface S from the interior $V_{2}$ and taking the cross products of eqs.~\eqref{eq:representationE2} and \eqref{eq:representationH2} with the surface normal vector $\mathbf{\hat{n}}$, the second set of equations is obtained:
\begin{equation}
\frac{1}{2}\mathbf{M}_{2} - \mathcal{K}^{2}_{em}\mathbf{M}_{2} -
\mathcal{K}^{2}_{ee}\mathbf{J}_{2} = \mathbf{0} 
\label{eq:EFIE2}
\end{equation}
\begin{equation}
\frac{1}{2}\mathbf{J}_{2} + \mathcal{K}^{2}_{mm}\mathbf{M}_{2} +
\mathcal{K}^{2}_{me}\mathbf{J}_{2} = \mathbf{0}
\label{eq:MFIE2}
\end{equation}
with the integral operator $\mathcal{K}^{2}_{\alpha\beta}(\alpha ,\beta \in \{ e,m\})$ defined the same way as in \eqref{eq:operatorK1}, except the superscript ``2" now indicates the uniaxially anisotropic interior region $V_{2}$. 

Note that due to the tangential continuity conditions of the fields across the boundary, we must have that
\begin{equation}
    \mathbf{J} = \mathbf{J}_{1} = -\mathbf{J}_{2}, \quad \mathbf{M} = \mathbf{M}_{1} = -\mathbf{M}_{2},
\end{equation}
which leaves two remaining unknowns $\mathbf{J}$ and $\mathbf{M}$ and four equations. As is commonly done, the four equations can be reduced to two via linear combination:
\begin{equation}
\begin{aligned}
    \alpha_{1} \eqref{eq:EFIE1}
    &+ \alpha_{2} \eqref{eq:EFIE2} \\
    \beta_{1} \eqref{eq:MFIE1}
    &+ \beta_{2} \eqref{eq:MFIE2}
    \label{eq:BIE}
\end{aligned}
\end{equation}
which is the system of integral equations that is used in our formulation. After the equivalent surface densities $\mathbf{J}$ and $\mathbf{M}$ have been solved, the total fields outside and inside the uniaxial object can be determined anywhere by evaluating the representation formulas \eqref{eq:representationE1}, \eqref{eq:representationH1} and \eqref{eq:representationE2}, \eqref{eq:representationH2} respectively. The specific choice of coefficients $\alpha$ and $\beta$, and the explicit form of each dyadic Green's function will be explained in the next section. 
\section{Dyadic Green's functions for uniaxial anisotropic media}
The interior region $V_{2}$ in the formulation is filled with a uniaxially anisotropic dielectric, which can characterized by the relative permittivity tensor:
\begin{equation}
\overline{\boldsymbol{\epsilon}} = \epsilon_{\perp}\overline{\mathbf{I}}+ (\epsilon_{\parallel}-\epsilon_{\perp})\hat{\mathbf{c}}\hat{\mathbf{c}}   
\end{equation}
where $\hat{\mathbf{c}}$ is a unit vector parallel to the distinguished axis, $\epsilon_{\parallel}$ is the relative permittivity along the direction of $\hat{\mathbf{c}}$, $\epsilon_{\perp}$ is the relative permittivity along the directions perpendicular to $\hat{\mathbf{c}}$ and $\overline{\mathbf{I}}$ represents the unit dyadic. It has been shown in \cite{mumcu2008surface,weiglhofer1990dyadic} that closed-form expressions exist for the dyadic Green's functions for this type of material, which we reproduce here for completeness:
\begin{equation}
    \begin{aligned}
        \overline{\mathbf{G}}_{ee}^{2} &= \frac{i\omega\mu_{0}}{4\pi}
        \bigg\{ \frac{\nabla\nabla}{\mathrm{k}_{\perp}^{2}}\frac{e^{i\mathrm{k}_{\perp}\mathrm{R_{e}}}}{\mathrm{R_{e}}} + \epsilon_{\parallel} \frac{e^{i\mathrm{k}_{\perp}\mathrm{R_{e}}}}{\mathrm{R_{e}}} \overline{\boldsymbol{\epsilon}}^{-1} \\ 
        &-\left[\frac{\epsilon_{\parallel}e^{i\mathrm{k}_{\perp}\mathrm{R_{e}}}}{\epsilon_{\perp}\mathrm{R_{e}}}-\frac{e^{i\mathrm{k}_{\perp}\mathrm{R}}}{\mathrm{R}}\right]\left[\frac{(\mathbf{R}\times\hat{\mathbf{c}})(\mathbf{R}\times\hat{\mathbf{c}})}{(\mathbf{R}\times\hat{\mathbf{c}})^{2}}\right] \\
        & - \left[ \frac{\epsilon_{\parallel}-\epsilon_{\perp}}{\epsilon_{\perp}}\frac{e^{i\mathrm{k}_{\perp}(\mathrm{R_{e}+R})/2}}{\mathrm{R_{e}+R}}
    \frac{\sin{(\mathrm{k}_{\perp}(\mathrm{R_{e}-R})/2)}}{(\mathrm{k}_{\perp}(\mathrm{R_{e}-R})/2)} \right] \\
    & \times \left[ \overline{\mathbf{I}}-\hat{\mathbf{c}}\hat{\mathbf{c}}-2\frac{(\mathbf{R}\times\hat{\mathbf{c}})(\mathbf{R}\times\hat{\mathbf{c}})}{(\mathbf{R}\times\hat{\mathbf{c}})^{2}} \right]
    \bigg\} 
    \end{aligned}
    \label{eq:Gee2}
\end{equation}

\begin{equation}
\begin{aligned}
          \overline{\mathbf{G}}_{mm}^{2} &= \frac{i\omega\epsilon_{0}}{4\pi}
        \bigg\{ \frac{\nabla\nabla}{\mathrm{k}_{0}^{2}}\frac{e^{i\mathrm{k}_{\perp}\mathrm{R}}}{\mathrm{R}} + \epsilon_{\perp} \frac{e^{i\mathrm{k}_{\perp}\mathrm{R}}}{\mathrm{R}} \overline{\mathbf{I}} \\ 
        &+\left[\frac{\epsilon_{\parallel}e^{i\mathrm{k}_{\perp}\mathrm{R_{e}}}}{\mathrm{R_{e}}}-\frac{\epsilon_{\perp}e^{i\mathrm{k}_{\perp}\mathrm{R}}}{\mathrm{R}}\right]\left[\frac{(\mathbf{R}\times\hat{\mathbf{c}})(\mathbf{R}\times\hat{\mathbf{c}})}{(\mathbf{R}\times\hat{\mathbf{c}})^{2}}\right] \\
        & + \left[ (\epsilon_{\parallel}-\epsilon_{\perp})\frac{e^{i\mathrm{k}_{\perp}(\mathrm{R_{e}+R})/2}}{\mathrm{R_{e}+R}}
    \frac{\sin{(\mathrm{k}_{\perp}(\mathrm{R_{e}-R})/2)}}{(\mathrm{k}_{\perp}(\mathrm{R_{e}-R})/2)} \right] \\
    & \times \left[ \overline{\mathbf{I}}-\hat{\mathbf{c}}\hat{\mathbf{c}}-2\frac{(\mathbf{R}\times\hat{\mathbf{c}})(\mathbf{R}\times\hat{\mathbf{c}})}{(\mathbf{R}\times\hat{\mathbf{c}})^{2}} \right]
    \bigg\} 
    \end{aligned}
    \label{eq:Gmm2}
\end{equation}

\begin{equation}
\overline{\mathbf{G}}_{em}^{2} = \frac{i}{\omega\epsilon_{0}}
\overline{\boldsymbol{\epsilon}}^{-1} \cdot \nabla \times \overline{\mathbf{G}}_{mm}^{2}
    \label{eq:Gem2}
\end{equation}

\begin{equation}
\overline{\mathbf{G}}_{me}^{2} = 
\frac{1}{i\omega\mu_{0}}
 \nabla \times \overline{\mathbf{G}}_{ee}^{2}
    \label{eq:Gme2}
\end{equation}
where $\epsilon_{0}$ and $\mu_{0}$ are the permittivity and permeability of free space respectively, $\omega$ is the angular frequency of the incident field, $\mathrm{k_{0}} = \omega \sqrt{\epsilon_{0}\mu_{0}}$ is the wavenumber in free space, $\mathbf{R} = \mathbf{r} - \mathbf{r'}$ and $\mathrm{R} = |\mathbf{R}|$ are the relative position vector and the distance respectively from a source point to an observation point, $\overline{\boldsymbol{\epsilon}}^{-1} = \epsilon_{\perp}^{-1}\overline{\mathbf{I}}+ (\epsilon_{\parallel}^{-1}-\epsilon_{\perp}^{-1})\hat{\mathbf{c}}\hat{\mathbf{c}} $ is the inverse of $\overline{\boldsymbol{\epsilon}}$ and $\mathrm{R_{e}}$, and $\mathrm{k}_{\perp}$ are given by:  
\begin{equation}
    \mathrm{R_{e}} = \sqrt{\epsilon_{\parallel}(\mathbf{R} \cdot \overline{\boldsymbol{\epsilon}}^{-1} \cdot \mathbf{R} )}, \quad \mathrm{k}_{\perp} =  \mathrm{k_{0}} \sqrt{\epsilon_{\perp}}
\end{equation}
Note that if the permittivity tensor is set to $\overline{\boldsymbol{\epsilon}} = \overline{\mathbf{I}}$, the above uniaxially anisotropic Green's functions simplify to the well-known isotropic dyadic Green's functions for free-space:
\begin{equation}
\overline{\mathbf{G}}_{ee}^{1} = \frac{i\omega\mu_{0}}{4\pi} \left[ \frac{\nabla \nabla }{\mathrm{k_{0}^{2}}} \frac{e^{i\mathrm{k_{0}}\mathrm{R}}}{\mathrm{R}}+\frac{e^{i\mathrm{k_{0}}\mathrm{R}}}{\mathrm{R}} \overline{\mathbf{I}} \right]
    \label{eq:Gee1}
\end{equation}

\begin{equation}
\overline{\mathbf{G}}_{mm}^{1} = 
\frac{i\omega\epsilon_{0}}{4\pi} \left[ \frac{\nabla \nabla }{\mathrm{k_{0}^{2}}} \frac{e^{i\mathrm{k_{0}}\mathrm{R}}}{\mathrm{R}}+\frac{e^{i\mathrm{k_{0}}\mathrm{R}}}{\mathrm{R}} \overline{\mathbf{I}} \right]
    \label{eq:Gmm1}
\end{equation}

\begin{equation}
\overline{\mathbf{G}}_{em}^{1} = \frac{i}{\omega\epsilon_{0}} \nabla \times \overline{\mathbf{G}}_{mm}^{1}
    \label{eq:Gem1}
\end{equation}

\begin{equation}
\overline{\mathbf{G}}_{me}^{1} = 
\frac{1}{i\omega\mu_{0}}
 \nabla \times \overline{\mathbf{G}}_{ee}^{1}
    \label{eq:Gme1}
\end{equation}
The linear combination coefficients in the integral equation system \eqref{eq:BIE} are chosen according to the N-M\"uller formulation to be: $\alpha_{1} = \epsilon_{r1} = 1, \alpha_{2} = \epsilon_{r2} = \epsilon_{\perp}, \beta_{1} = \mu_{r1} = \beta_{2} = \mu_{r2} =1$, which cancel the singularity of the hypersingular part of the $\overline{\mathbf{G}}_{ee}^{1}$ operator and result in a well-conditioned second-kind integral equation formulation~\cite{yla2005well}. The resulting integral equations can be represented in matrix form as follows:
\begin{multline}              \begin{bmatrix}
    \mathcal{K}^{1}_{em}-\epsilon_{\perp}\mathcal{K}^{2}_{em}+\frac{1+\epsilon_{\perp}}{2}\mathcal{I} &
    \mathcal{K}^{1}_{ee}-\epsilon_{\perp}\mathcal{K}^{2}_{ee}\\
    \mathcal{K}^{2}_{mm} - \mathcal{K}^{1}_{mm} &
    \mathcal{K}^{2}_{me} - \mathcal{K}^{1}_{me}+\mathcal{I}
    \end{bmatrix} 
    \begin{bmatrix} 
    \mathbf{M} \\
    \mathbf{J} 
    \end{bmatrix} \\
    =\begin{bmatrix} 
    -\mathbf{\hat{n}} \times \mathbf{E}^{\text{inc}} \\
    \mathbf{\hat{n}} \times \mathbf{H}^{\text{inc}}
    \end{bmatrix}
    \label{Nmuller}
\end{multline} 
where $\mathcal{I}$ is the identity operator and the expressions for dyadic Green's functions $\overline{\mathbf{G}}^{i}_{\alpha\beta}$ involved in each of the integral operators $\mathcal{K}^{i}_{\alpha\beta} (i \in \{1,2\}; \alpha ,\beta \in \{ e,m\})$ are given by \eqref{eq:Gee2}-- \eqref{eq:Gme2} and \eqref{eq:Gee1}-- \eqref{eq:Gme1}.

\section{Singularity Analysis of Integral Operators\label{sec:singularity}}
In order to evaluate the action of each of the integral operators $\mathcal{K}^{i}_{\alpha\beta}$ on the densities with high accuracy, care must be taken to analyze and properly handle the singular behavior of each operator. 
\subsection{Singularity of $\mathcal{K}^{2}_{ee}$}
At first glance, the $\mathcal{K}^{i}_{ee}$ and $\mathcal{K}^{i}_{mm} (i \in \{1,2\}) $ operators appear to both be hypersingular with $O(1/\mathrm{R}^3)$ singularities due to the $\nabla\nabla$ operator acting on a term with $O(1/\mathrm{R})$ singularity. However, vector identities can be utilized to transfer the one of the $\nabla$ operators to the density term and the other $\nabla$, which can be made to not depend on the source integration coordinate, can be pulled outside of the integral\footnote{Note: Moving the gradient ($\nabla$) outside the integral is not strictly necessary when using the M\"uller formulation since its coefficients are designed to cancel the singularity.}. For example, taking the $\mathcal{K}^{2}_{ee}$
operator with a target point approaching the surface from the inside,
\begin{equation}
\begin{aligned}
    \mathcal{K}^{2}_{ee}\mathbf{J} &= \mathbf{\hat{n}}(\mathbf{r}) \times \int_{S} \overline{\mathbf{G}}_{ee}^{2}(\mathbf{r},\mathbf{r}')\cdot\mathbf{J}(\mathbf{r}')d\sigma(\mathbf{r'}) \biggr|_{\mathbf{r} \in S} \\
    & = \mathbf{\hat{n}}(\mathbf{r}) \times \int_{S} \overline{\mathbf{G}}_{ee}^{2}(\mathbf{r},\mathbf{r}')\cdot\mathbf{J}(\mathbf{r}')d\sigma(\mathbf{r'}) \biggr|_{\mathbf{r} \to \mathbf{r}^{-}} \\
    &= \frac{i\omega\mu_{0}}{4\pi} \mathbf{\hat{n}}(\mathbf{r}) \times \int_{S} \bigg\{ \frac{\nabla\nabla}{\mathrm{k}_{\perp}^{2}}\frac{e^{i\mathrm{k}_{\perp}\mathrm{R_{e}}}}{\mathrm{R_{e}}} +\overline{\mathbf{D}}\bigg\}\cdot\mathbf{J}(\mathbf{r}')d\sigma(\mathbf{r'}) \biggr|_{\mathbf{r} \to \mathbf{r}^{-}} \\
    & = \frac{i\omega\mu_{0}}{4\pi} \mathbf{\hat{n}}(\mathbf{r}) \times \bigg\{\frac{1}{\mathrm{k}_{\perp}^{2}} \nabla \int_{S}\nabla\frac{e^{i\mathrm{k}_{\perp}\mathrm{R_{e}}}}{\mathrm{R_{e}}}  \cdot\mathbf{J}(\mathbf{r}')d\sigma(\mathbf{r'})\\
    &+\int_{S} \overline{\mathbf{D}} \cdot \mathbf{J}(\mathbf{r}')d\sigma(\mathbf{r'})\bigg\}
    \biggr|_{\mathbf{r} \to \mathbf{r}^{-}} \\
    & = \frac{i\omega\mu_{0}}{4\pi} \mathbf{\hat{n}}(\mathbf{r}) \times \bigg\{\frac{1}{\mathrm{k}_{\perp}^{2}} \nabla \int_{S}\frac{e^{i\mathrm{k}_{\perp}\mathrm{R_{e}}}}{\mathrm{R_{e}}}  \nabla'_{s}\cdot\mathbf{J}(\mathbf{r}')d\sigma(\mathbf{r'}) \\
    &+\int_{S} \overline{\mathbf{D}} \cdot \mathbf{J}(\mathbf{r}')d\sigma(\mathbf{r'})\bigg\} \biggr|_{\mathbf{r} \to \mathbf{r}^{-}}
    \label{eq:Kee2J}
\end{aligned}
\end{equation}
where 
\begin{equation}
    \begin{aligned}
    \overline{\mathbf{D}} &= \epsilon_{\parallel} \frac{e^{i\mathrm{k}_{\perp}\mathrm{R_{e}}}}{\mathrm{R_{e}}} \overline{\boldsymbol{\epsilon}}^{-1} -\left[\frac{\epsilon_{\parallel}e^{i\mathrm{k}_{\perp}\mathrm{R_{e}}}}{\epsilon_{\perp}\mathrm{R_{e}}}-\frac{e^{i\mathrm{k}_{\perp}\mathrm{R}}}{\mathrm{R}}\right]\left[\frac{(\mathbf{R}\times\hat{\mathbf{c}})(\mathbf{R}\times\hat{\mathbf{c}})}{(\mathbf{R}\times\hat{\mathbf{c}})^{2}}\right] \\
        & - \left[ \frac{\epsilon_{\parallel}-\epsilon_{\perp}}{\epsilon_{\perp}}\frac{e^{i\mathrm{k}_{\perp}(\mathrm{R_{e}+R})/2}}{\mathrm{R_{e}+R}}
    \frac{\sin{(\mathrm{k}_{\perp}(\mathrm{R_{e}-R})/2)}}{(\mathrm{k}_{\perp}(\mathrm{R_{e}-R})/2)} \right] \\
    & \times \left[ \overline{\mathbf{I}}-\hat{\mathbf{c}}\hat{\mathbf{c}}-2\frac{(\mathbf{R}\times\hat{\mathbf{c}})(\mathbf{R}\times\hat{\mathbf{c}})}{(\mathbf{R}\times\hat{\mathbf{c}})^{2}} \right]
    \end{aligned}
    \label{eq:Ddyadic}
\end{equation}
and $\mathbf{r} \to \mathbf{r}^{-}$ indicates that operator is evaluated for a target point that is approaching $\mathbf{r} \in S$ along $-\mathbf{\hat{n}}$ from $V_{2}$.  Since the kernels of both integrals, $\overline{\mathbf{D}}$ and $e^{i\mathrm{k}_{\perp}\mathrm{R_{e}}}/\mathrm{R_{e}}$, have $O(1/\mathrm{R})$ singularity, the integral operator $\mathcal{K}^{2}_{ee}$ in this form is weakly singular. $\mathcal{K}^{2}_{mm}, \mathcal{K}^{1}_{ee}$ and $ \mathcal{K}^{1}_{mm}$ can also be readily transformed into weakly singular operators by following the same procedure as $\mathcal{K}^{2}_{ee}$.
\subsection{Singularity of $\mathcal{K}^{2}_{me}$}
The action of the $\nabla \times$ operator on weakly singular kernels with $O(1/\mathrm{R})$ singularities makes the dyadic Green's functions of the $\mathcal{K}^{i}_{me}$ and $\mathcal{K}^{i}_{em} (i \in \{1,2\})$ operators strongly singular with $O(1/\mathrm{R}^{2})$ type singularity. Nevertheless, these operators can also be manipulated to become weakly singular when acting on densities by applying vector identities. For example, consider the $\mathcal{K}^{2}_{me}$ acting on $\mathbf{J}$, with the target point $\mathbf{r}$ approaching the surface from the inside as before,
\begin{equation}
    \begin{aligned}
        \mathcal{K}^{2}_{me}\mathbf{J} &= \mathbf{\hat{n}}(\mathbf{r}) \times \int_{S} \overline{\mathbf{G}}_{me}^{2}(\mathbf{r},\mathbf{r}')\cdot\mathbf{J}(\mathbf{r}')d\sigma(\mathbf{r'}) \biggr|_{\mathbf{r} \in S} \\
        & = \mathbf{\hat{n}}(\mathbf{r}) \times \int_{S} \overline{\mathbf{G}}_{me}^{2}(\mathbf{r},\mathbf{r}')\cdot\mathbf{J}(\mathbf{r}')d\sigma(\mathbf{r'}) \biggr|_{\mathbf{r} \to \mathbf{r}^{-}} + \frac{1}{2}\mathbf{J}\\
        & = \mathbf{\hat{n}}(\mathbf{r}) \times \int_{S} \frac{1}{i\omega\mu_{0}}
 \nabla \times \overline{\mathbf{G}}_{ee}^{2}\cdot\mathbf{J}(\mathbf{r}')d\sigma(\mathbf{r'})\biggr|_{\mathbf{r} \to \mathbf{r}^{-}} + \frac{1}{2}\mathbf{J}\\
 & = \frac{1}{4\pi} \mathbf{\hat{n}}(\mathbf{r}) \times \bigg\{ \int_{S}
 \nabla \times \frac{\nabla\nabla}{\mathrm{k}_{\perp}^{2}}\frac{e^{i\mathrm{k}_{\perp}\mathrm{R_{e}}}}{\mathrm{R_{e}}} \cdot\mathbf{J}(\mathbf{r}')d\sigma(\mathbf{r'}) \biggr|_{\mathbf{r} \to \mathbf{r}^{-}} \\
 & + \int_{S} \nabla \times \overline{\mathbf{D}}(\mathbf{r},\mathbf{r}') \cdot\mathbf{J}(\mathbf{r}')d\sigma(\mathbf{r'}) \biggr|_{\mathbf{r} \to \mathbf{r}^{-}} \bigg\} + \frac{1}{2}\mathbf{J} \\
 & = \frac{1}{4\pi} \mathbf{\hat{n}}(\mathbf{r})\times  \nabla \times \int_{S}
  \overline{\mathbf{D}}(\mathbf{r},\mathbf{r}') \cdot\mathbf{J}(\mathbf{r}')d\sigma(\mathbf{r'}) \biggr|_{\mathbf{r} \to \mathbf{r}^{-}} + \frac{1}{2}\mathbf{J}
  \label{eq:Kme2J}
    \end{aligned}
\end{equation}
where $\overline{\mathbf{D}}$ is given in \eqref{eq:Ddyadic} and the second equality follows from the jump condition. Note that the $\nabla\nabla$ term can be removed since $\nabla \times \nabla \equiv 0$. It can be seen that the kernel inside the integral ($\overline{\mathbf{D}}(\mathbf{r},\mathbf{r}')$) is now weakly singular since the curl operation has been factored out of the integral. The same procedure can be used to also transform $\mathcal{K}^{2}_{em}$,  $\mathcal{K}^{1}_{me}$ and $\mathcal{K}^{1}_{em}$ into weakly singular forms.

These operators in their weakly singular form can now be discretized with high-order accuracy using the Chebyshev-based Nystr\"om method that was first introduced in~\cite{hu2021chebyshev} for perfect conductors and isotropic dielectric materials. The following section briefly reviews the key points of the Chebyshev method and discusses our adaptation and application of it to the present anisotropic formulation.

\section{Evaluation of Action of Integral Operators $\mathcal{K}^{i}_{\alpha\beta}$ using Chebyshev expansion based Method \label{sec:actionevaluation}}
According to the analysis in section \ref{sec:singularity}, two types of weakly-singular integrals as well as their gradient and curl need to be evaluated to compute the action of the integral operators $\mathcal{K}^{2}_{ee}$ and $\mathcal{K}^{2}_{me}$ on the current density $\mathbf{J}$:
\begin{equation}
\begin{aligned}
    \phi(\mathbf{r}) &= \int_{S}\frac{e^{i\mathrm{k}_{\perp}\mathrm{R_{e}}}}{\mathrm{R_{e}}}  \nabla'_{s}\cdot\mathbf{J}(\mathbf{r}')d\sigma(\mathbf{r'}),\quad  \mathbf{\hat{n}}(\mathbf{r}) \times \nabla \phi(\mathbf{r}) |_{\mathbf{r} = \mathbf{r}^{-}} \\
    \mathbf{A}(\mathbf{r}) &= \int_{S} \overline{\mathbf{D}}(\mathbf{r},\mathbf{r}') \cdot \mathbf{J}(\mathbf{r}')d\sigma(\mathbf{r'}),\quad \mathbf{\hat{n}}(\mathbf{r}) \times \nabla \times \mathbf{A}(\mathbf{r})|_{\mathbf{r} = \mathbf{r}^{-}}
    \end{aligned}
\end{equation}
where $\phi(\mathbf{r})$ and $\mathbf{A}(\mathbf{r})$ are scalar and vector functions of the target point $\mathbf{r}$ respectively and $\overline{\mathbf{D}}$ is defined in~\eqref{eq:Ddyadic}. Note that we focus on the operators acting on $\mathbf{J}$ since the same procedure can be used to discretize the $K^2_{mm}$ and $K^2_{em}$ operators which act on $\mathbf{M}$.
\subsection{Evaluation of $\phi(\mathbf{r})$ and $\mathbf{A}(\mathbf{r})$ \label{subsec:A}}
In order to compute $\phi(\mathbf{r})$ and $\mathbf{A}(\mathbf{r})$, the whole surface $S$ is split into $M$ non-overlapping curvilinear quadrilateral patches $S_{p}, p = 1,2,...,M$. A parametric mapping is defined from the unit square $[-1,1]\times[-1,1]$ in UV space to each surface $S_p$ in Cartesian coordinates. Specifically, we introduce parameterization $\mathbf{r} = \mathbf{r}^p(u,v) = \left(x^p(u,v),y^p(u,v),z^p(u,v)\right)$ for patch $S_{p}$. The tangential covariant basis vectors and normal vectors on $S_{p}$ can then be defined as
\begin{equation}
    \mathbf{a}^p_u = \frac{\partial\mathbf{r}^p(u,v)}{\partial u},
    \; 
    \mathbf{a}^p_v = \frac{\partial\mathbf{r}^p(u,v)}{\partial v},
    \; 
    \mathbf{\hat{n}}^p = \frac{\mathbf{a}^p_u \times \mathbf{a}^p_v}{||\mathbf{a}^p_u \times \mathbf{a}^p_v||}.
\end{equation}
The tangential electric current density vector $\mathbf{J}$ on the surface $S_{p}$ can be expanded in terms of the local tangential coordinate basis as
\begin{align}
   \mathbf{J}^p (u,v)
    = J^{p,u}(u,v)\mathbf{a}^p_u(u,v) + J^{p,v}(u,v)\mathbf{a}^p_v(u,v)
    \label{Jpdef} 
\end{align}
where $\mathbf{J}^p (u,v) \equiv \mathbf{J} (\mathbf{r}^p(u,v)) $, $J^{p,u}$ and $J^{p,v}$ are the contravariant components of the surface current density $\mathbf{J}$. For sufficiently smooth surface geometries, $J^{p,u}$ and $J^{p,v}$ are smooth functions of $u$ and $v$ and can be approximated with spectral convergence by using Chebyshev polynomials as
\begin{equation}
    J^{p,a} = \sum_{m=0}^{N^p_v-1} \sum_{n=0}^{N^p_u-1} \gamma^{p,a}_{n,m}T_n(u)T_m(v),\quad \text{for $a=u,v$}
\label{eq:Jdiscr}
\end{equation}
where the Chebyshev coefficients $\gamma^{p,a}_{n,m}$ can be computed from the values of $J^{p,a}$ on $S_p$ at the Chebyshev nodes, which is where the discretized set of unknowns are located, by using the discrete orthogonality property of Chebyshev polynomials:
\begin{equation}
    \gamma^{p,a}_{n,m} = \frac{\alpha_n\alpha_m}{N^p_uN^p_v}\sum_{k=0}^{N^p_v-1} \sum_{l=0}^{N^p_u-1} J^{p,a}(u_l,v_k)T_n(u_l)T_m(v_k),
    \label{eq:Chebycoef}
\end{equation}
After the Chebyshev coefficients are obtained from the density values on Chebyshev nodes, we are able to compute the density values $J^{p,a}(u,v)$ for arbitrary $(u,v)$ by interpolating via \eqref{eq:Jdiscr}, and the Cartesian components $J^{p}_{i}(u,v)$ can be computed by taking dot product of Cartesian basis vectors $\mathbf{e}_{i} (i=x,y,z)$ and $\mathbf{J}^p (u,v)$. Thus, $\phi(\mathbf{r})$ and the $i$-th Cartesian component of the integral  $\mathbf{A}(\mathbf{r})$ can be represented as
\begin{equation}
\begin{aligned}
     \phi(\mathbf{r})  &= \sum_{p=1}^{M}\int_{S_{p}}\frac{e^{i\mathrm{k}_{\perp}\mathrm{R_{e}}}}{\mathrm{R_{e}}}  \nabla'_{s}\cdot\mathbf{J}(\mathbf{r}')d\sigma(\mathbf{r'}) \\
    & = \sum_{p=1}^{M}\int_{-1}^{1}\int_{-1}^{1}\frac{e^{i\mathrm{k}_{\perp}\mathrm{R_{e}}}}{\mathrm{R_{e}}}  \biggr(\frac{\partial(\sqrt{|G^p|}J^{p,u})}{\partial u}\\
    & + \frac{\partial(\sqrt{|G^p|}J^{p,v})}{\partial v}\biggr) dudv 
     \end{aligned}
\end{equation}
\begin{equation}
    \begin{aligned}
    A_{i}(\mathbf{r}) &= \sum_{p=1}^{M}\int_{S_{p}} \mathbf{e}_{i} \cdot
    \overline{\mathbf{D}}(\mathbf{r},\mathbf{r}') \cdot \mathbf{J}(\mathbf{r}')d\sigma(\mathbf{r'}) \\
    & = \sum_{p=1}^{M}\int_{-1}^{1}\int_{-1}^{1}(D_{ix}J^{p}_{x}+D_{iy}J^{p}_{y}+D_{iz}J^{p}_{z})\sqrt{|G^p|}dudv
    \end{aligned}
\end{equation}
where $D_{ij} = D_{ij}(\mathbf{r},\mathbf{r}^p(u,v)) (i,j = x,y,z)$ is the Cartesian component of the dyadic $ \overline{\mathbf{D}}(\mathbf{r},\mathbf{r}')$, $J^{p}_{j} = J^{p}_{j}(u,v)$ is the Cartesian component of current density $\mathbf{J}$ and $\sqrt{|G^p|} = \sqrt{|G^p(u,v)|}$ is the surface element Jacobian on the source patch $S_{p}$. If the target point $\mathbf{r}$ is far away from $S_{p}$, the kernels $D_{ij}$ and $e^{i\mathrm{k}_{\perp}\mathrm{R_{e}}}/\mathrm{R_{e}}$ are smooth and Fejer's first quadrature rule can be used directly on the discrete densities at the Chebyshev nodes to evaluate the integrals numerically with high-order accuracy. When the target point $\mathbf{r}$ is on the source patch $S_{p}$ itself or nearby, the integrals become singular or nearly singular and require special treatment. Since the density on each patch can be expanded in terms of a Chebyshev polynomial basis via \eqref{eq:Jdiscr}, the action of these integrals on the density $\mathbf{J}$ can be computed by first precomputing their action on each Chebyshev basis polynomial, followed by multiplying the resulting values against the expanded Chebyshev coefficients of the density and accumulating over all $n$ and $m$ indices. Since all of the kernels involved have been manipulated to be weakly singular, we adopt the change of variables proposed in \cite{hu2021chebyshev,bruno2020chebyshev} \cite[Sec. 3.5]{colton1998inverse} to regularize the integrals by annihilating the singularity with the surface Jacobian, allowing the precomputations to be computed with very high accuracy using a standard Fejer quadrature rule. The Chebyshev discretization and singular integration approaches for the Nystr\"om method are described in depth in \cite{hu2021chebyshev}.     
\subsection{Evaluation of $\mathbf{\hat{n}}(\mathbf{r}) \times \nabla \phi(\mathbf{r}) |_{\mathbf{r} = \mathbf{r}^{-}}$ \label{subsec:B}}
In view of the surface representation in terms of non-overlapping patches, for a target point $\mathbf{r}$ on $p$th patch $S_{p}$, we first expand the $\nabla$ operator in the local coordinate frame as
\begin{equation}
    \nabla = \mathbf{a}^{p,u} \frac{\partial}{\partial u} + \mathbf{a}^{p,v} \frac{\partial}{\partial v} + \mathbf{\hat{n}}^p \frac{\partial}{\partial \mathbf{\hat{n}}^p}
    \label{eq:delexpand}
\end{equation}
where $\mathbf{a}^{p,u}$ and $\mathbf{a}^{p,v}$ are contravariant basis vectors that satisfy the orthogonality relation
\begin{equation}
    \mathbf{a}^{p,a} \cdot \mathbf{a}^p_b = \begin{cases}
    1 & a=b\\
    0 & a\neq b
    \end{cases}.
\end{equation}
The operator can then be expanded as
\begin{equation}
\begin{aligned}
    & \mathbf{\hat{n}}(\mathbf{r}) \times \nabla \phi(\mathbf{r})
    |_{\mathbf{r} \to \mathbf{r}^{-}} \\
    &= \mathbf{\hat{n}}^p \times (\mathbf{a}^{p,u} \frac{\partial \phi}{\partial u} + \mathbf{a}^{p,v} \frac{\partial \phi}{\partial v} +
    \mathbf{\hat{n}}^p \frac{\partial \phi }{\partial \mathbf{\hat{n}}^p}) \biggr|_{\mathbf{r} \to \mathbf{r}^{-}} \\ 
    & =  \frac{\partial \phi}{\partial u}\biggr|_{\mathbf{r} \to \mathbf{r}^{-}} \mathbf{\hat{n}}^p \times \mathbf{a}^{p,u} + \frac{\partial \phi}{\partial v}\biggr|_{\mathbf{r} \to \mathbf{r}^{-}} \mathbf{\hat{n}}^p \times \mathbf{a}^{p,v} \\
    & =  \frac{\partial \phi}{\partial u}\biggr|_{\mathbf{r} \in S_{p}} \mathbf{\hat{n}}^p \times \mathbf{a}^{p,u} + \frac{\partial \phi}{\partial v}\biggr|_{\mathbf{r} \in S_{p}} \mathbf{\hat{n}}^p \times \mathbf{a}^{p,v}
\end{aligned}
\label{eq:ncrossGradPhi}
\end{equation}
Note that the third equality follows from the fact that $\phi(\mathbf{r})$ has continuous tangential derivatives across the surface without any jump condition. As in section \ref{subsec:A}, $\phi(\mathbf{r})$ is first computed at each Chebyshev node $(u_l,v_k)$ on $S_{p}$ and then expanded with a Chebyshev transform as
\begin{equation}
    \phi(\mathbf{r}^p(u,v)) = \sum_{m=0}^{N^p_v-1} \sum_{n=0}^{N^p_u-1} \zeta^{p}_{n,m}T_n(u)T_m(v)
\end{equation}
where $\zeta^{p}_{n,m}$ are the Chebyshev coefficients obtained by using \eqref{eq:Chebycoef} and replacing $J^{p,a}(u_l,v_k)$ with $\phi(\mathbf{r}^p(u_l,v_k))$. The partial derivatives with respect to $u$ and $v$ can then be readily computed by taking the derivatives of Chebyshev polynomials $T_n(u)$ and $T_m(v)$ respectively as
\begin{equation}
\begin{aligned}
    \frac{\partial \phi}{\partial u}(\mathbf{r}^p(u,v)) &= \sum_{m=0}^{N^p_v-1} \sum_{n=0}^{N^p_u-1} \zeta^{p}_{n,m}T_n'(u)T_m(v) \\
    \frac{\partial \phi}{\partial v}(\mathbf{r}^p(u,v)) &= \sum_{m=0}^{N^p_v-1} \sum_{n=0}^{N^p_u-1} \zeta^{p}_{n,m}T_n(u)T_m'(v)
\end{aligned}
\end{equation}
for all target points $\mathbf{r} = \mathbf{r}^p(u,v) \in S_{p}$ and $\mathbf{\hat{n}}(\mathbf{r}) \times \nabla \phi(\mathbf{r}) |_{\mathbf{r} = \mathbf{r}^{-}}$ can then be computed by substituting into expansion \eqref{eq:ncrossGradPhi}. 
\subsection{Evaluation of $\mathbf{\hat{n}}(\mathbf{r}) \times \nabla \times \mathbf{A}(\mathbf{r})|_{\mathbf{r} = \mathbf{r}^{-}}$}
By using the same expansion for $\nabla$ operator as in \eqref{eq:delexpand}, we can expand this operator as
\begin{equation}
    \begin{aligned}
        &\mathbf{\hat{n}}(\mathbf{r}) \times \nabla \times \mathbf{A}(\mathbf{r})|_{\mathbf{r} \to \mathbf{r}^{-}} \\
        & =\mathbf{\hat{n}}^p \times (\mathbf{a}^{p,u} \times \frac{\partial \mathbf{A}}{\partial u} + \mathbf{a}^{p,v} \times \frac{\partial \mathbf{A}}{\partial v} + \mathbf{\hat{n}}^p \times \frac{\partial \mathbf{A}}{\partial \mathbf{\hat{n}}^p}) \biggr|_{\mathbf{r} \to \mathbf{r}^{-}} \\
        & = \mathbf{\hat{n}}^p \times (\mathbf{a}^{p,u} \times \frac{\partial \mathbf{A}}{\partial u} \biggr|_{\mathbf{r} \to \mathbf{r}^{-}}) + \mathbf{\hat{n}}^p \times (\mathbf{a}^{p,v} \times \frac{\partial \mathbf{A}}{\partial v} \biggr|_{\mathbf{r} \to \mathbf{r}^{-}}) \\
        & + \mathbf{\hat{n}}^p \times (\mathbf{\hat{n}}^p \times \frac{\partial \mathbf{A}}{\partial \mathbf{\hat{n}}^p} \biggr|_{\mathbf{r} \to \mathbf{r}^{-}}) \\
        & = \mathbf{a}^{p,u}(\mathbf{\hat{n}}^p \cdot \frac{\partial \mathbf{A}}{\partial u} \biggr|_{\mathbf{r} \in S_{p}}) + \mathbf{a}^{p,v}(\mathbf{\hat{n}}^p \cdot \frac{\partial \mathbf{A}}{\partial v} \biggr|_{\mathbf{r} \in S_{p}}) - \frac{\partial \mathbf{A}}{\partial \mathbf{\hat{n}}^p} \biggr|_{\mathbf{r} \to \mathbf{r}^{-}} \\
        &+ \mathbf{\hat{n}}^p(\mathbf{\hat{n}}^p \cdot \frac{\partial \mathbf{A}}{\partial \mathbf{\hat{n}}^p} \biggr|_{\mathbf{r} \to \mathbf{r}^{-}})
    \end{aligned} 
\end{equation}
where the tangential derivatives for each Cartesian component of $\mathbf{A}$, $\frac{\partial \mathbf{A}}{\partial u}$ and $\frac{\partial \mathbf{A}}{\partial v}$, on $S_{p}$ can be evaluated in the same way as $\frac{\partial \phi}{\partial u}$ and $\frac{\partial \phi}{\partial v}$ in section \ref{subsec:B}.

According to the limit definition of the directional derivative, the normal derivative of each Cartesian component $i$ of $\mathbf{A}$, $\frac{\partial \mathbf{A}_i}{\partial \mathbf{\hat{n}}^p} |_{\mathbf{r} \to \mathbf{r}^{-}}$, can be written as:
\begin{equation}
    \frac{\partial \mathrm{A}_{i}}{\partial \mathbf{\hat{n}}^p} \biggr|_{\mathbf{r} \to \mathbf{r}^{-}} = \lim_{\delta \to 0^{+}} \frac{\mathrm{A}_{i}(\mathbf{r})-\mathrm{A}_{i}(\mathbf{r}-\delta \mathbf{\hat{n}}^p)}{\delta} \quad i = x,y,z
\end{equation}
The normal derivative can be transformed into a derivative of a univariate function by definining auxiliary function, $g(\delta) = \mathrm{A}_{i}(\mathbf{r}+\delta \mathbf{\hat{n}}^p)$:
\begin{equation}
    \begin{aligned}
    \frac{\partial \mathrm{A}_{i}}{\partial \mathbf{\hat{n}}^p} \biggr|_{\mathbf{r} = \mathbf{r}^{-}} &= \lim_{\delta \to 0^{+}} \frac{\mathrm{A}_{i}(\mathbf{r})-\mathrm{A}_{i}(\mathbf{r}-\delta \mathbf{\hat{n}}^p)}{\delta} \\
    & = \lim_{\delta \to 0^{+}} \frac{g(0)-g(-\delta)}{\delta} = g'_{-}(0)
    \end{aligned}
\end{equation}
In order to approximate the derivative $g'_{-}(0)$ numerically with high accuracy without requiring very close off-surface evaluation, we use the following backward difference approximation:
\begin{equation}
\begin{aligned}
    \frac{\partial \mathrm{A}_{i}}{\partial \mathbf{\hat{n}}^p} \biggr|_{\mathbf{r} = \mathbf{r}^{-}} &= g'_{-}(0) \approx \frac{3g(0)-4g(-\delta)+g(-2\delta)}{\delta} \\
    &= \frac{3\mathrm{A}_{i}(\mathbf{r})-4\mathrm{A}_{i}(\mathbf{r}-\delta \mathbf{\hat{n}}^p)+\mathrm{A}_{i}(\mathbf{r}-2\delta \mathbf{\hat{n}}^p)}{\delta} 
\end{aligned}
\label{eq:finite_difference}
\end{equation}
which results in second order accuracy $\mathcal{O}(\delta^2)$ as $\delta \to 0^{+}$. Note that the weakly singular integrals $\mathrm{A}_{i}(\mathbf{r})$, $\mathrm{A}_{i}(\mathbf{r}-\delta \mathbf{\hat{n}}^p)$ and $\mathrm{A}_{i}(\mathbf{r}-2\delta \mathbf{\hat{n}}^p)$ in the numerator can be evaluated with high accuracy using the rectangular-singular integration method discussed in section \ref{subsec:A}.

After the two weakly singular integrals $\phi(\mathbf{r})$ and $\mathbf{A}(\mathbf{r})$ and their gradient and curl have been evaluated respectively, $\mathcal{K}^{2}_{ee} \mathbf{J}$ and $\mathcal{K}^{2}_{me} \mathbf{J}$ can be obtained by substituting into \eqref{eq:Kee2J} and \eqref{eq:Kme2J}. The same approach can be used to compute the actions of the other integral operators required since the kernels of $\mathcal{K}^{2}_{mm} \mathbf{M}$, $\mathcal{K}^{1}_{ee} \mathbf{J}$ and $\mathcal{K}^{1}_{mm} \mathbf{M}$ are similar to that of $\mathcal{K}^{2}_{ee} \mathbf{J}$ and the kernels of $\mathcal{K}^{2}_{em} \mathbf{M}$, $\mathcal{K}^{1}_{me} \mathbf{J}$ and $\mathcal{K}^{1}_{em} \mathbf{M}$ are similar to that of $\mathcal{K}^{2}_{me} \mathbf{J}$ as discussed in section \ref{sec:singularity}. Therefore, the LHS of the whole system \eqref{Nmuller} can be evaluated for an arbitrary target point $\mathbf{r} \in S$. As is done in a typical Nystr\"om method, the operators are evaluated at the same targets points as the unknowns; i.e., at the Chebyshev nodes on each patch, and each equation is tested with the two tangential contravariant basis vectors. This results in a full-rank linear system with the same number of equations as unknowns, which can readily be solved using a suitable linear solver of choice. In this work, we use GMRES to solve the discretized systems iteratively.

\section{Numerical Results\label{sec:results}}
We first study the convergence of the forward map with respect to the number of Chebyshev nodes per side of the patch: $N=N^{p}_{u}=N^{p}_{v}$ of the forward map. This can be done numerically by applying the whole system \eqref{Nmuller} operator, which includes the actions of all the integral operators, on reference current densities $\mathbf{J}$ and $\mathbf{M}$ on a sphere and comparing against an analytical Mie series solution~\cite{geng2004mie}. Following this, we present several examples demonstrating scattering from a uniaxially anisotropic dielectric sphere and cube to highlight the high order accuracy which can be achieved with our method. We also solve a scattering example from a 3D NURBS model generated by a commercial CAD software to demonstrate the ability of our method to handle objects with complicated geometrical features and curvature. Finally, we apply our method to a silicon nanophotonic phase-shifter waveguiding structure and compare the results against a commercial FDTD solver to showcase the potential of our method for simulating nanophotonic devices with high accuracy.

\subsection{Forward Map Convergence} \label{subsec:6-A}
\begin{figure}[ht]
\centering
\subfloat[][]{
\includegraphics[width=80mm]{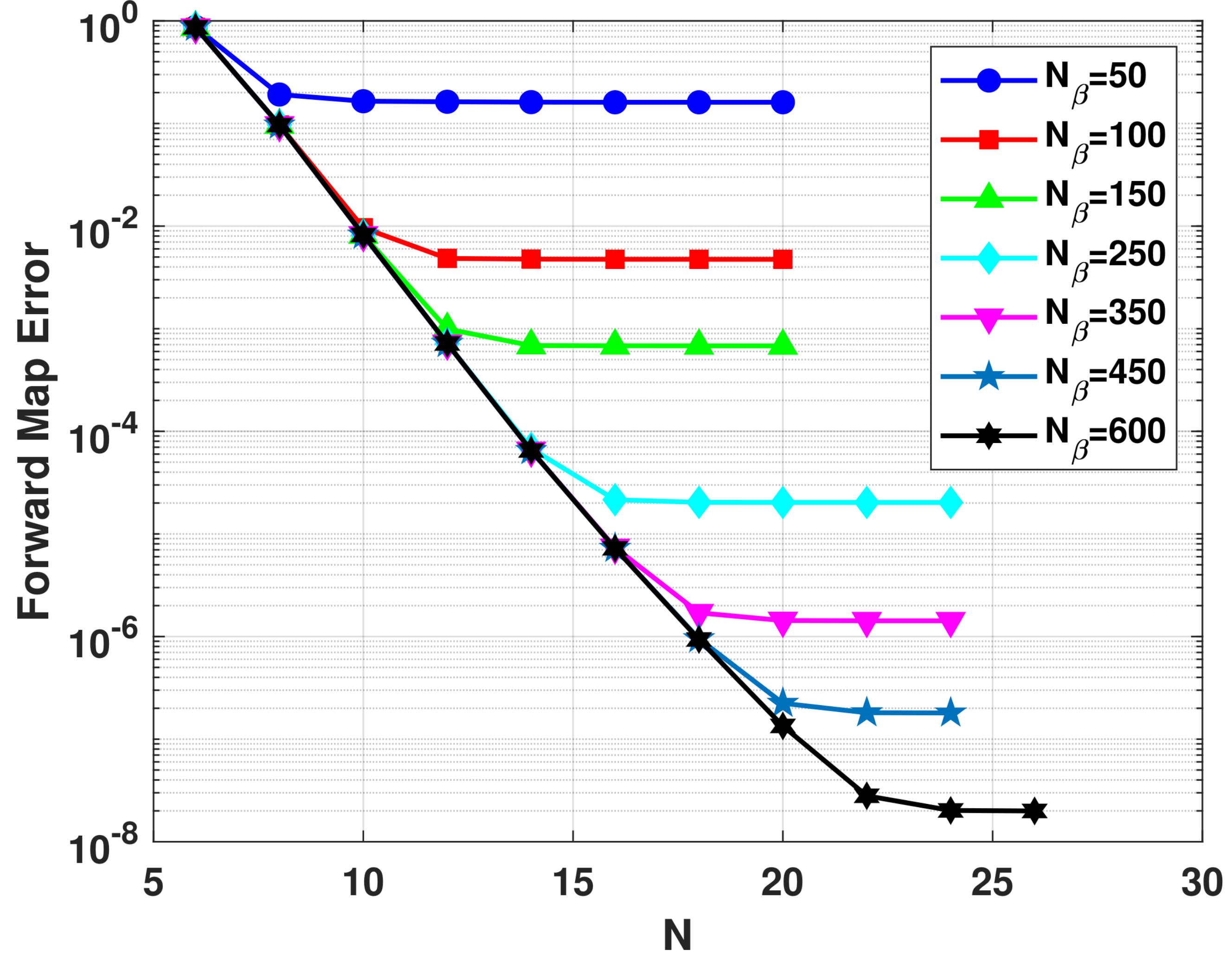}
}\\
\subfloat[][]{
\includegraphics[width=80mm]{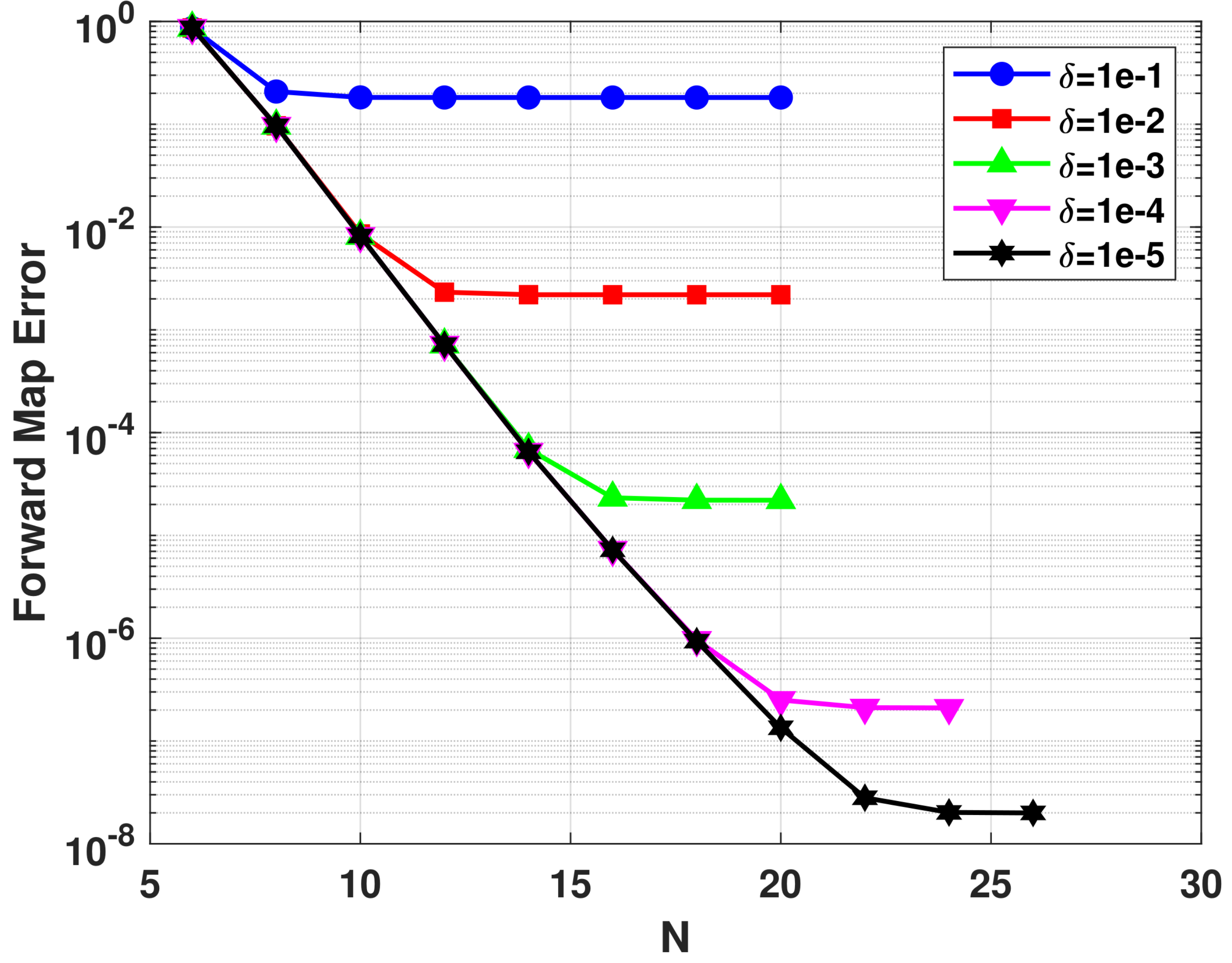}
}
\caption{(a) Forward mapping error with respect to $N$ for different $N_\beta$ on a uniaxially anisotropic dielectric sphere ($D=2\lambda_{0}$) with finite difference step size $\delta = 10^{-5}$. (b) Forward mapping error with respect to $N$ for different step size $\delta$ on the same sphere with $N_\beta = 600$.}
\label{fig:fwd_map_error}
\end{figure}

We evaluate convergence of the forward map (application of the integral operator to a prescribed density) on a uniaxially anisotropic dielectric sphere with diameter $D$ = $2\lambda_{0}$, anisotropic permittivity $\epsilon_{\perp} = 2$, $\epsilon_{\parallel} = 3$ and distinguished axis $\hat{\mathbf{c}} = (0,0,1)$. Fig.~\ref{fig:fwd_map_error}(a) and (b) plot the forward mapping error versus $N$ for increasing $N_{\beta}$ and decreasing finite difference step size $\delta$ in \eqref{eq:finite_difference} respectively. Note that a sufficiently small $\delta = 10^{-5}$ is used for the plot versus $N_{\beta}$ and a sufficienly large $N_{\beta} = 600$ is used for the plot versus $\delta$ such that the convergence is dominated by the parameter that is under consideration in each plot. An analytical Mie series solution for scattering from a uniaxially anisotropic dielectric sphere due to an incident plane wave~\cite{geng2004mie} is used for the reference densities. As expected and discussed in Section~\ref{sec:actionevaluation}, both the $\delta$ and $N_{\beta}$ parameters affect the overall accuracy significantly and should be chosen judiciously according to the desired overall solution accuracy. 
\subsection{Uniaxially Anisotropic Sphere}
Next we investigate solving the full scattering problem for the same sphere considered in section \ref{subsec:6-A}. The electric field of the incident plane wave is given by  $\mathbf{E}^\text{inc} = \mathbf{e}_{x}e^{ik_{0}z}$. To verify the correctness and accuracy of our results, the result of our solver is compared with the analytical Mie series solution~\cite{geng2004mie}.

\begin{figure}[htp]
\centering
\subfloat[][]{
    \includegraphics[width=35mm]{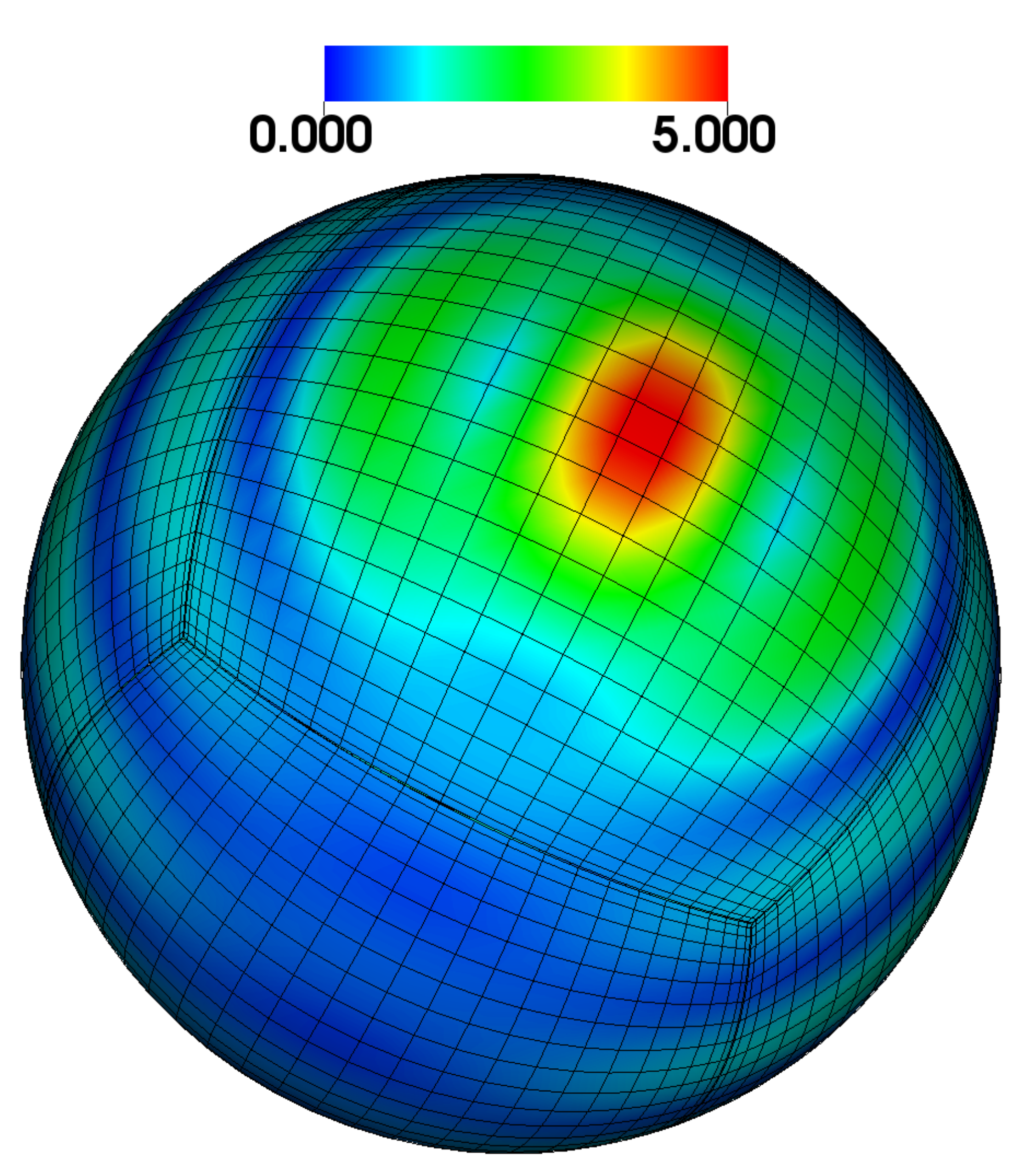}
}
\subfloat[][]{
    \includegraphics[width=35mm]{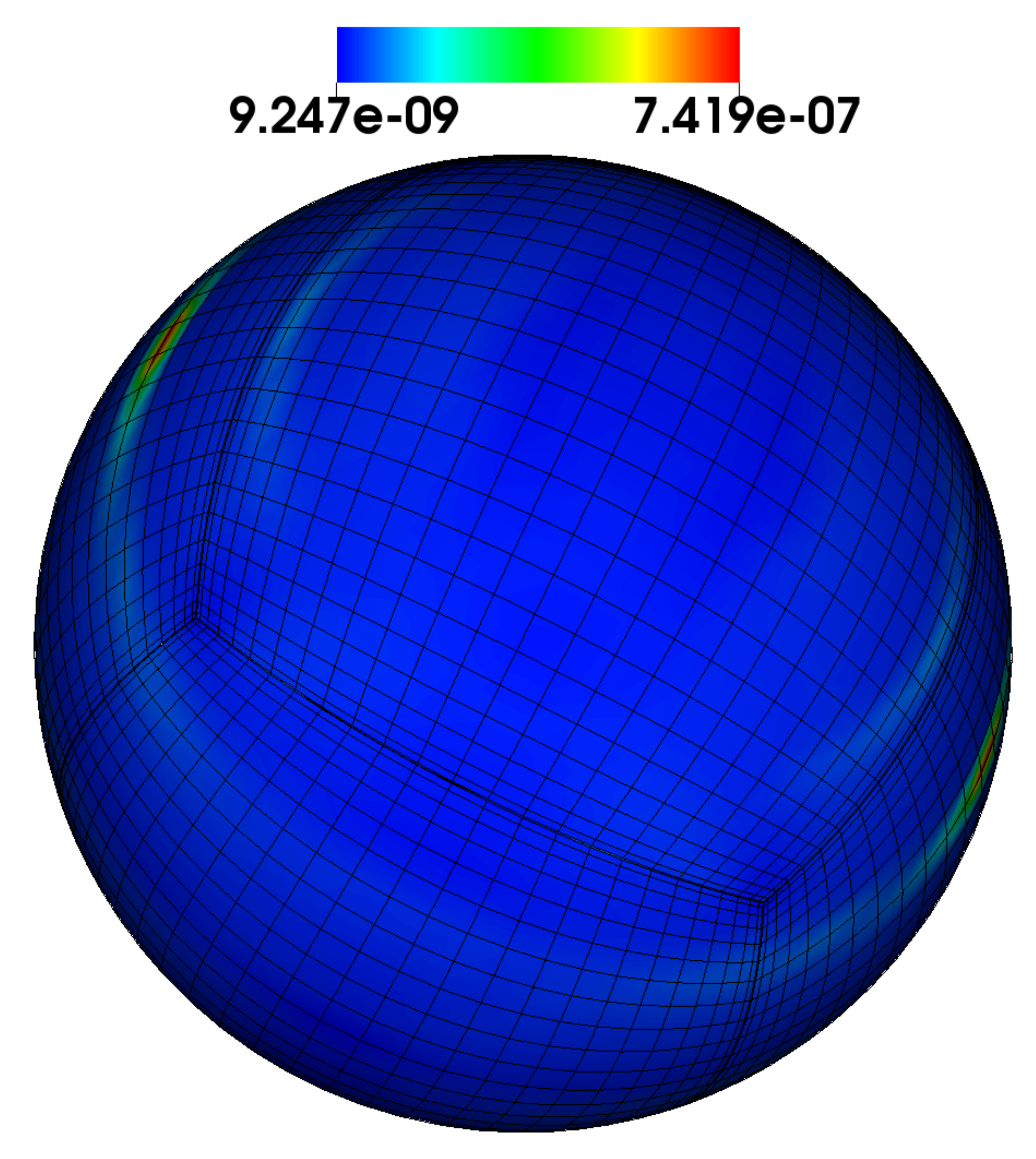}
} 
\\
\subfloat[][]{
    \includegraphics[width=35mm]{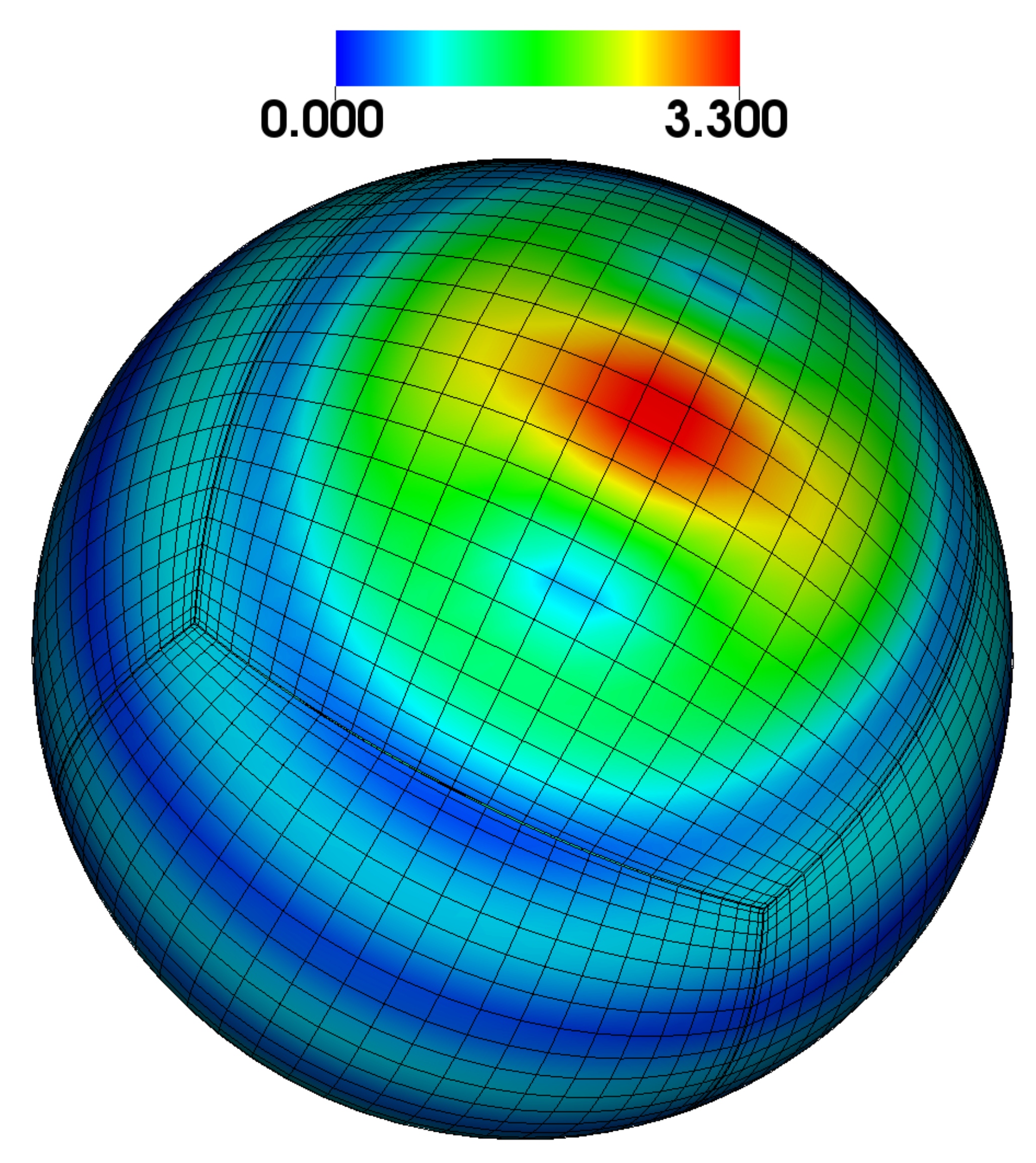}
}
\subfloat[][]{
\includegraphics[width=35mm]{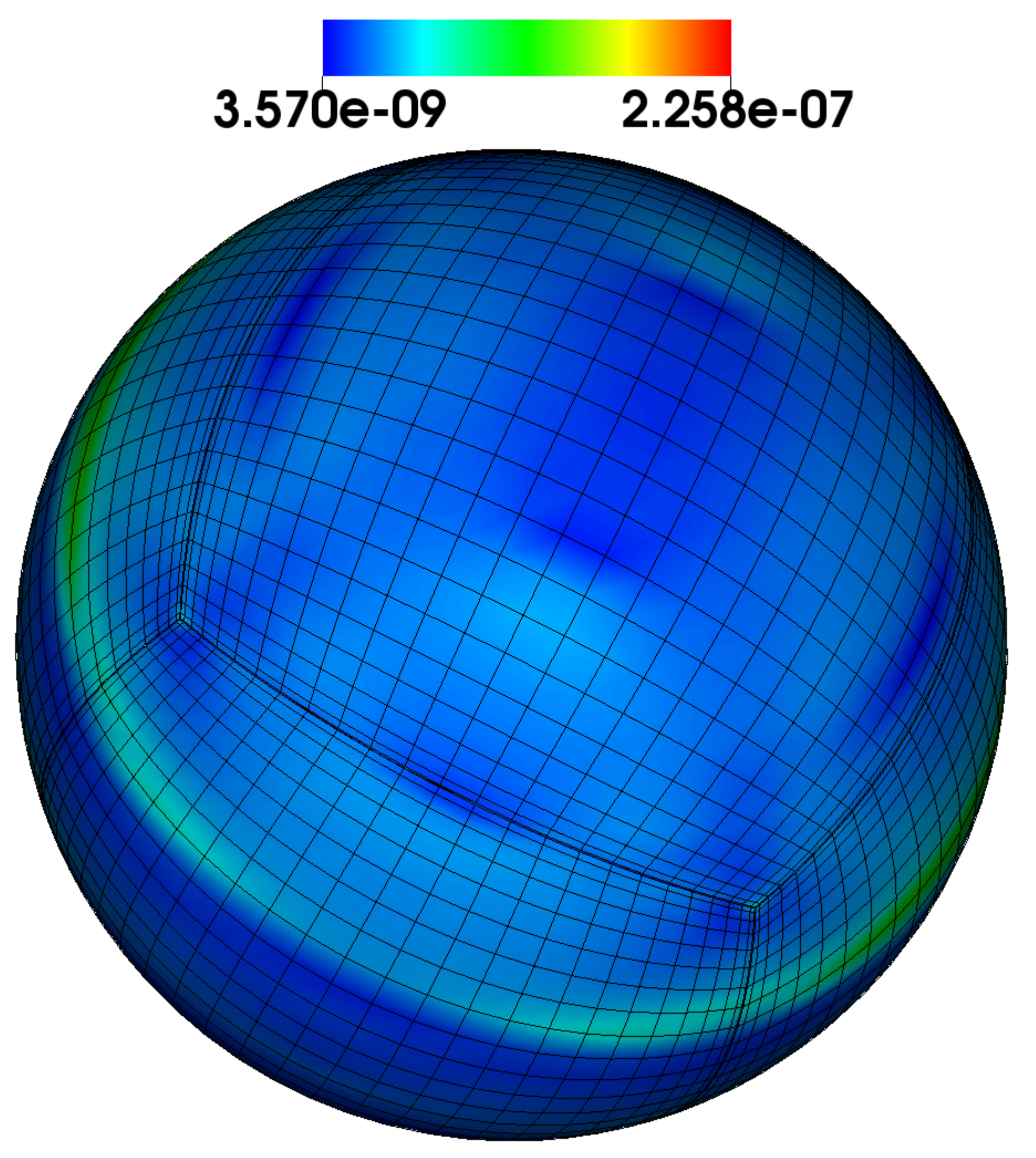}
}
\\
\subfloat[][]{
    \includegraphics[width=70mm]{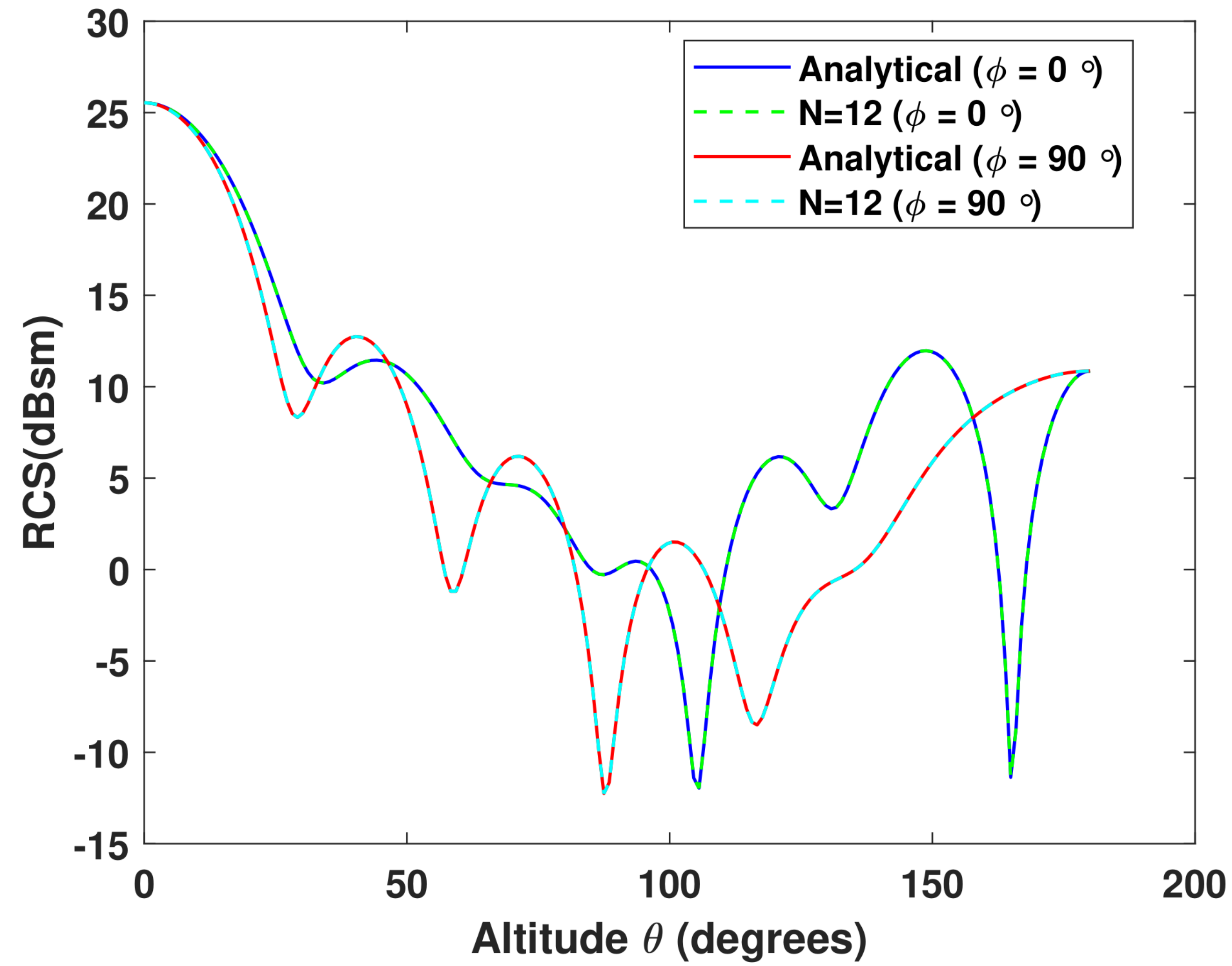}
    }
    \\
\subfloat[][]{
    \includegraphics[width=70mm]{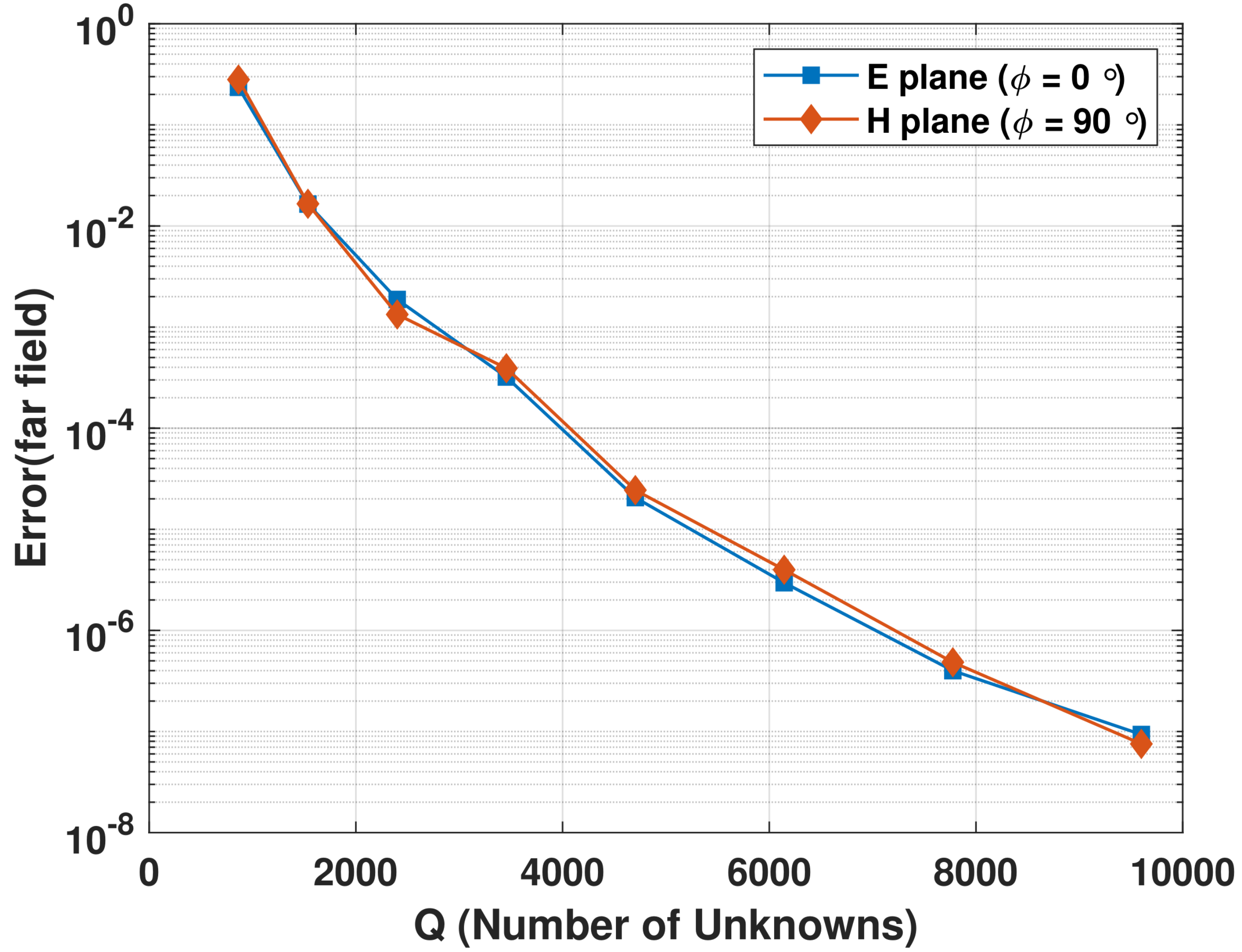}
    } 
\caption{(a) Magnitude of surface electric current density distribution $|\mathbf{J}|$ on a uniaxially anisotropic dielectric sphere ($D = 2\lambda_{0}$) induced by incident plane wave. (b) Error of surface $|\mathbf{J}|$ distribution. Max error: $7.4\times10^{-7}$. (c) Magnitude of surface magnetic current density distribution $|\mathbf{M}|$. (d) Error of surface $|\mathbf{M}|$ distribution. Max error: $2.3 \times 10^{-7}$. (e) RCS at E plane ($\phi= 0^{\circ}$) and H plane ($\phi= 90^{\circ}$) comparing a discretization of $N= 12$ with the exact solution. (f) Far-field relative error for both E and H planes with respect to the exact Mie series solution.}\label{fig:Sphere} 
\end{figure}

Fig.~\ref{fig:Sphere}(a) and (c) show the magnitudes of electric and magnetic surface current densities, $|\mathbf{J}|$ and $|\mathbf{M}|$, on the sphere for $N=24$. In Fig.~\ref{fig:Sphere}(b) and (d), we plot the associated error of each density on the surface with respect to the analytical solution. Fig.~\ref{fig:Sphere}(e) compares the RCS for both the E plane ($\phi=0^{\circ}$) and H plane ($\phi=90^{\circ}$) computed by using a discretization of $12\times 12$ points per patch versus the analytical solution. As can be seen, the results from the solver are indistinguishable from the analytical solution. Finally, Fig.~\ref{fig:Sphere}(f) plots the corresponding relative error in the RCS solution in both planes with respect to the analytical solution versus $N$ (number of points per side of each patch), demonstrating the solver's high-order convergence.

\subsection{Uniaxially Anisotropic Cube}
We also consider scattering from a uniaxially aniostropic dielectric cube with $1\lambda_{0}$ edge length, anisotropic permittivity $\epsilon_{\perp} = 3$, $\epsilon_{\parallel} = 5$ and distinguished axis $\hat{\mathbf{c}} = (\frac{1}{2},\frac{1}{2},\frac{\sqrt{2}}{2})$. The same plane wave incident field is used as the previous example and the surface of the cube is made up of $6$ patches. Since we are not aware of an analytical solution for this structure, we also compared the result of our solver with a highly refined solution ($N=40$) as well as with a solution obtained from a commercial Finite Element (FEM) simulation software.
\begin{figure}[htb]
    \centering
    \subfloat[][]{
    \includegraphics[width=40mm]{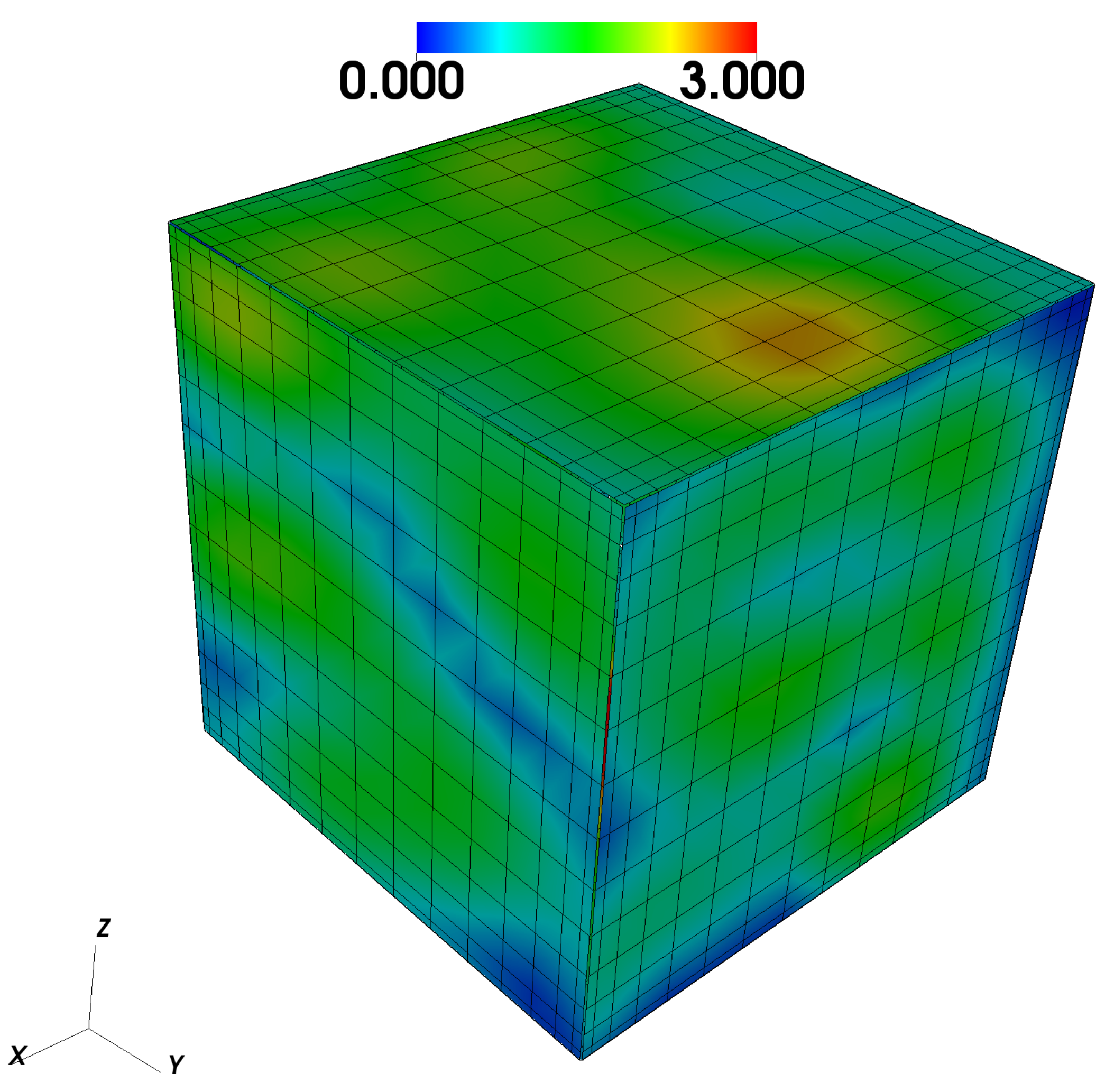}
    } 
    \subfloat[][]{
    \includegraphics[width=40mm]{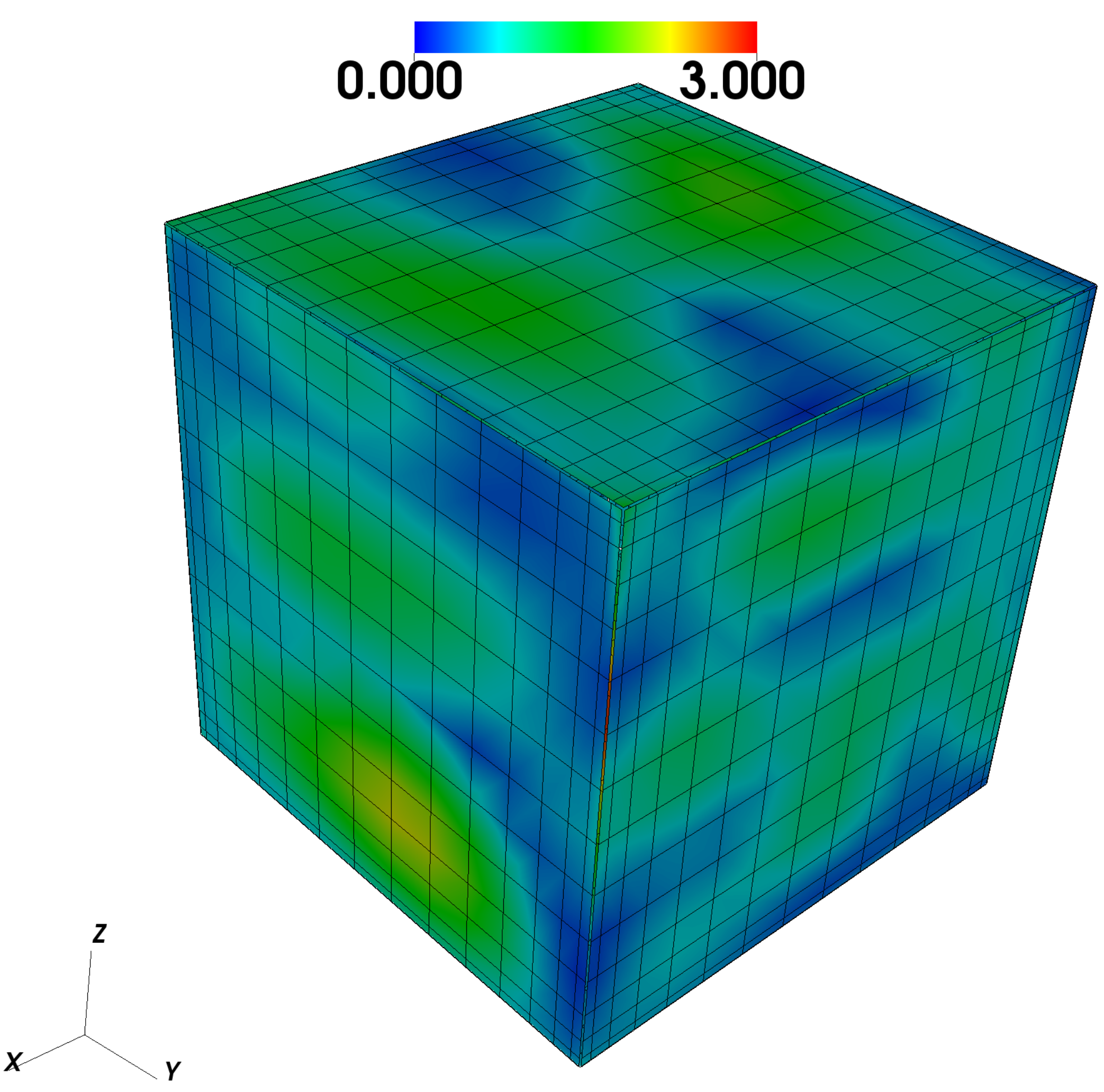}
    } \\
    \subfloat[][]{
    \includegraphics[width=70mm]{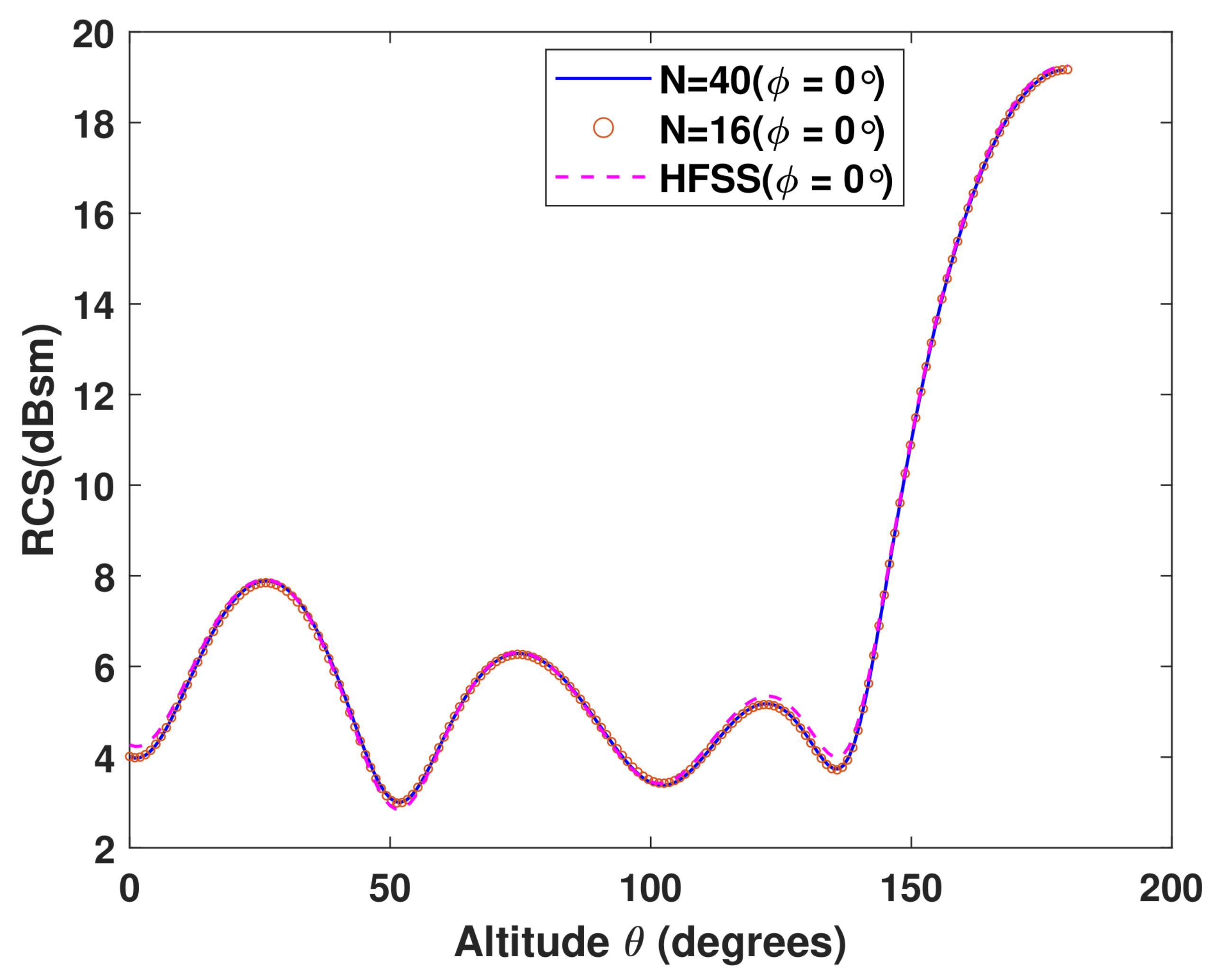}
    } 
    \caption{(a)  Magnitude of surface electric current density distribution $|\mathbf{J}|$ on a $1\lambda_{0}$ edge length uniaxially anisotropic dielectric cube. (b) Magnitude of surface magnetic current density distribution $|\mathbf{M}|$ on the same cube. (c) RCS at E plane ($\phi=0^{\circ}$) from a coarse discretization $N=16$, a refined discretization $N=40$m and the  commercial FEM solver, Ansys HFSS.}
    \label{fig:Cube}
\end{figure}

Fig.~\ref{fig:Cube}(a) and (b) show the magnitudes of the surface electric and magnetic current densities, $|\mathbf{J}|$ and $|\mathbf{M}|$, for $N=16$. Fig.~\ref{fig:Cube}(c) compares the RCS at the E plane ($\phi=0^{\circ}$) obtained by using a coarse discretization $N=16$, a highly refined discretization $N=40$, and the commercial FEM solver Ansys HFSS~\cite{ansys}. As can be seen, the results from $N=16$ and $N=40$ are completely overlapped with each other, demonstrating that the solver has already converged for a relatively coarse discretization, despite the known challenges with objects that have sharp edges and corners that often plague Mawxell solvers. A maximum deviation less than 0.3dB between our result and the FEM solver result is observed, which further validates the correctness and effectiveness of our solver for scatters with sharp edges. Note that no particular edge refinement strategy was used in this example, although a similar approach as the edge change of variables used in ~\cite{hu2021chebyshev} could be applied to improve the convergence further.

\subsection{Hummingbird 3D NURBS CAD Model}
We also compute the fields scattered by a hummingbird composed of a uniaxially anisotropic dielectric material. The hummingbird geometry used is a 3D NURBS CAD model that is available freely online \cite{hummingbird}. The same incident excitation and permittivity tensor settings are used as in the sphere example. The hummingbird is sized such that it has a total length of 4.3 wavelengths and a wingspan of 6.5 wavelengths. This geometry consists of 311 curvilinear quadrilateral patches that were generated by the commerical CAD software Rhino~\cite{rhino_cite}.

\begin{figure}[htb]
\centering
\subfloat[][]{
\includegraphics[width=60mm]{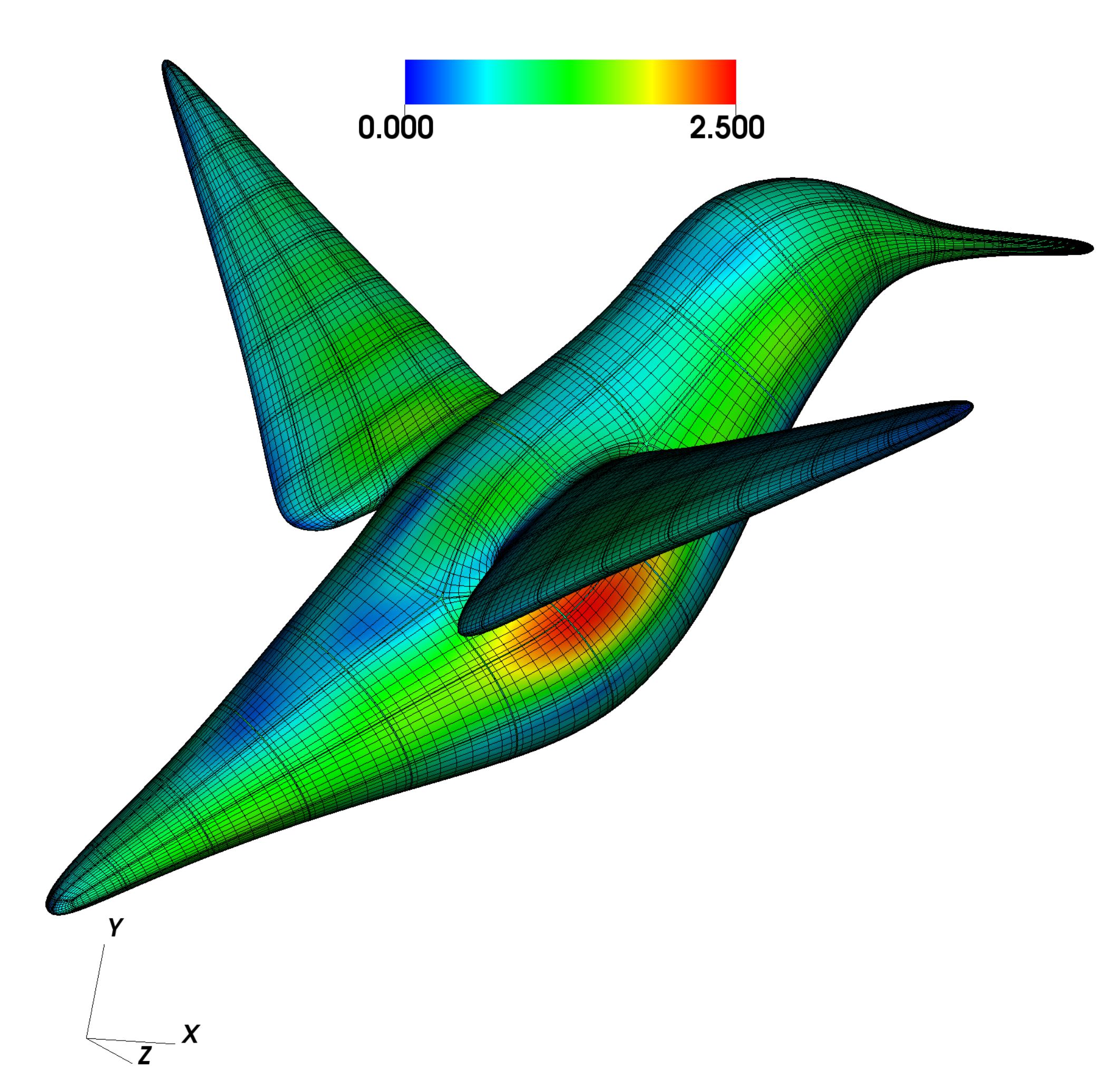}
}\\
\subfloat[][]{
\includegraphics[width=70mm]{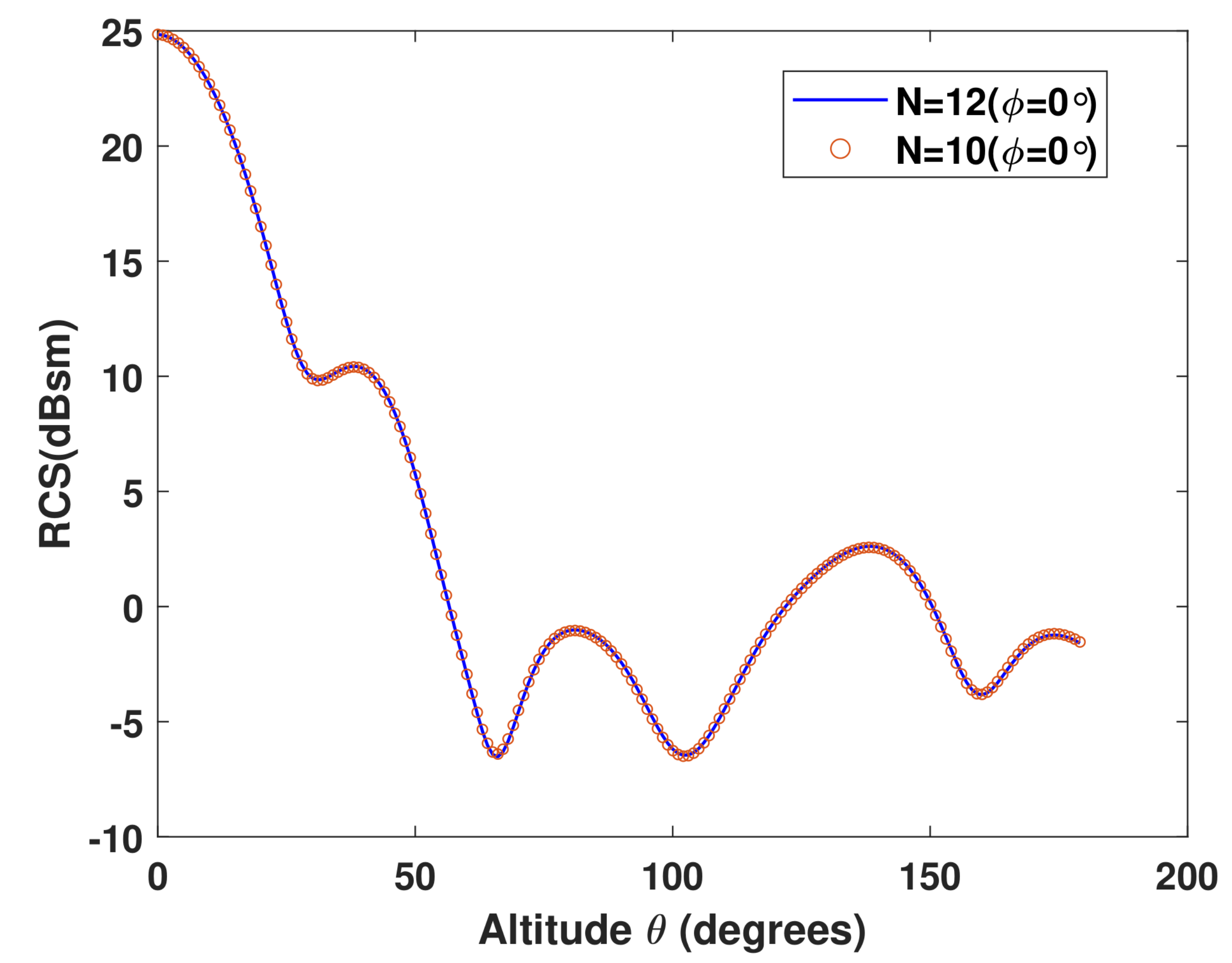}
} 
\caption{(a) Magnitude of surface magnetic current density $|\mathbf{M}|$ on a uniaxially anistropic dielectric hummingbird CAD model with surface composed of 311 patches. (b) RCS at E plane ($\phi=0^{\circ}$) from discretizations of $N=10$ and $N=12$ .}
\label{fig:CAD_bird}
\end{figure}
We plot the magnitude of surface magnetic current density $|\mathbf{M}|$ in Fig.~\ref{fig:CAD_bird}(a) and RCS versus $\theta$ at $\phi= 0^{\circ}$ for two discretizations $N=10$ and $N=12$ in Fig.~\ref{fig:CAD_bird}(b). Although this model contains sharp corners at the beak, tail, and wing tips that could be challenging to simulate accurately, very close agreement is observed for the RCS patterns resulting from the two discretizations. 

\subsection{Silicon Photonic Phase Shifter}
We conclude our numerical results section with one final example of a silicon-based nanophotonic phase-shifter embedded in a liquid crystal background medium. This is a simplified design inspired by~\cite{pfeifle2012silicon} and consists of two parallel rectangular silicon waveguides embedded within a uniaxially anisotropic liquid crystal cladding. The orientation of the distinguished axis $\hat{\mathbf{c}}$ of the liquid crystal media can be electrically controlled by an external voltage. By altering the amplitude of this voltage, the distinguished axis is rotated, causing the permittivity experienced by the dominant field component to change and leading to a different corresponding propagation constant. This changes the phase-shift experienced by light propagating in the fundamental mode of the waveguide over a certain distance as discussed in~\cite{pfeifle2012silicon}. 

In our example, the width and the height of the rectangular cross sections of both waveguides are $0.24\mu m$ and $0.22\mu m$ respectively, and the spacing between the two silicon rods is $0.12\mu m$. The anisotropic permittivity of the liquid crystal cladding is set to be $\epsilon_{\perp} = 2.3409$, $\epsilon_{\parallel} = 2.9241$, the distinguished axis $\hat{\mathbf{c}}$ is set to either $\hat{\mathbf{x}}$ or $\hat{\mathbf{z}}$, and the silicon waveguide has permittivity $\epsilon_{\mathrm{Si}} = 12.11$. We use an electric dipole polarized along $(1,0,0)$ direction with unit amplitude and $1.55\mu m$ free space wavelength placed at $(0,0,-1)$ as the source excitation. The Windowed Green Function (WGF) method is used to simulate the waveguides extending into infinity from both directions~\cite{bruno2017windowed,sideris2019ultrafast,garza2021boundary}.
\begin{figure}[h]
\centering
\subfloat[][]{
\includegraphics[width=70mm]{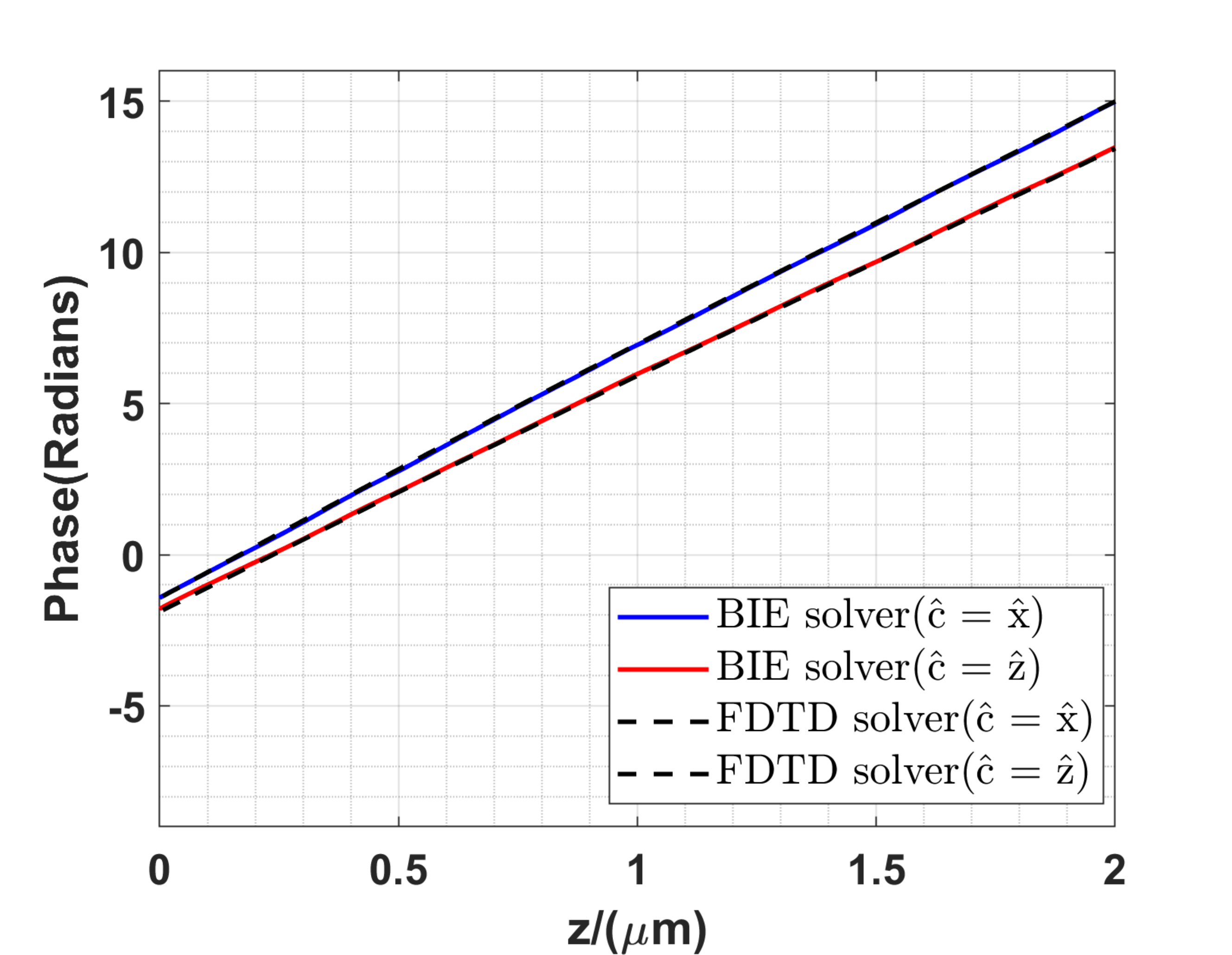}
}
\\
\subfloat[][]{
\includegraphics[width=70mm]{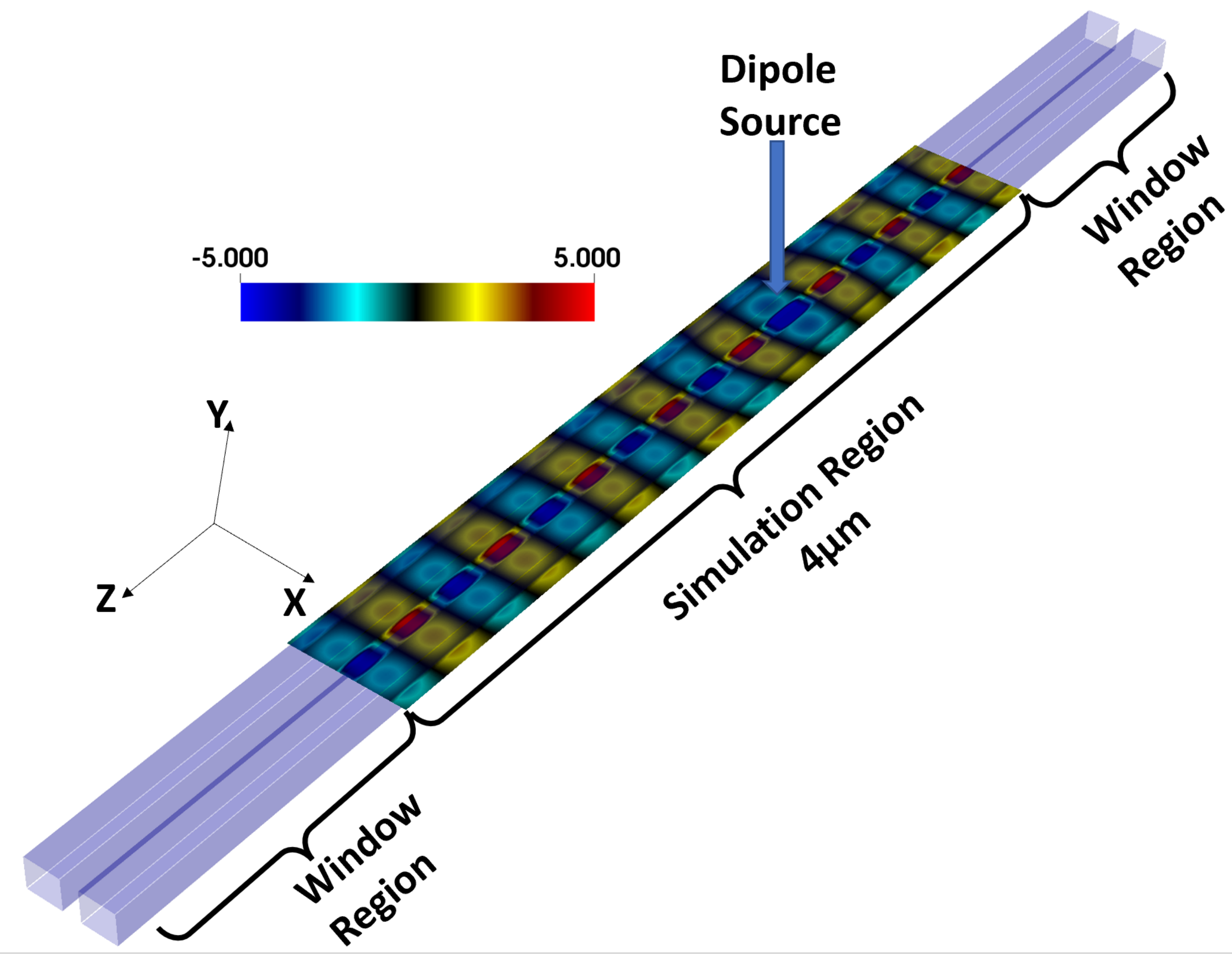}
} 
\caption{(a) Phase variation of $\mathrm{E}_{x}$ along the propagation direction for both $\hat{\mathbf{c}} = \hat{\mathbf{x}}$ and $\hat{\mathbf{c}} = \hat{\mathbf{z}}$. (b) Real part of $\mathrm{E}_{x}$ on the planar cross section $-0.4\mu m \leq x \leq 0.4\mu m,-2\mu m \leq z \leq 2\mu m$ }
\label{fig:phaseshifter}
\end{figure}

Fig.~\ref{fig:phaseshifter}(a) shows the phase variation of the dominant field component $\mathrm{E}_{x}$ along the propagation direction for both $\hat{\mathbf{c}} = \hat{\mathbf{x}}$ and $\hat{\mathbf{c}} = \hat{\mathbf{z}}$ obtained by using our solver as well as a commercial FDTD solver. The results of the two solvers match very closely with each other. As expected, due to the difference in the propagation constants of the propagating modes caused by rotating the distinguished axis of the liquid crystal cladding from $\hat{\mathbf{c}} = \hat{\mathbf{x}}$ to $\hat{\mathbf{c}} = \hat{\mathbf{z}}$, the slopes of the phase versus position for the two scenarios are different. The real part of the $\mathrm{E}_{x}$ field on the planar cross section $-0.4\mu m \leq x \leq 0.4\mu m,-2\mu m \leq z \leq 2\mu m$ is depicted in Fig.~\ref{fig:phaseshifter}(b), indicating single mode propagation along the waveguide.

\section{Conclusion}
We introduced a high-order accurate approach to solve the 3D Maxwell surface integral equation formulation for scattering from uniaxially anisotropic objects and media. Specifically, we utilized vector identities to represent the integral operators in terms of weakly singular integrals and their gradients and curls. A Chebyshev polynomial expansion based approach similar to the one used in our previous work for isotropic dielectric and metallic objects~\cite{hu2021chebyshev} is applied for discretizing and evaluating these operators numerically. The high accuracy of the method is verified by comparing the convergence of the solution for scattering from a uniaxial anisotropic dielectric sphere to an analytical solution. Other examples were also presented, including scattering from a uniaxially anisotropic cube, a 3D NURBS model generated by a commercial CAD software, and a silicon photonic phase-shifter embedded in a liquid crystal background medium, which demonstrate the effectiveness and versatility of the solver for handling many different scenarios. Future work includes using the solver to inverse design high-performance radio-frequency and nanophotonic devices using uniaxially anisotropic materials, such as liquid crystals, which can be dynamically reconfigured by switching their polarization states. 

\bibliographystyle{IEEEtran}
\bibliography{IEEEabrv,bibliography}

\end{document}